\documentclass[twocolumn,aps,pra,amsmath,floatfix]{revtex4}
\usepackage{graphicx}
\begin{document}

\title{Quasi-particle spectra of perovskites: 
Enhanced Coulomb correlations at surfaces}

\author{A. Liebsch}
\affiliation{Institut f\"ur Festk\"orperforschung, Forschungszentrum
             J\"ulich, 52425 J\"ulich, Germany}
\date{\today }

\begin{abstract}
Photoemission spectra of the perovskites  Ca$_x$Sr$_{1-x}$VO$_3$,
Ca$_x$La$_{1-x}$VO$_3$, and SrRuO$_3$ indicate that Coulomb correlations 
are more pronounced at the surface than in the bulk. To investigate this 
effect we use the dynamical mean field theory combined with the Quantum 
Monte Carlo technique and evaluate the multi-orbital self-energy. 
These systems exhibit different degrees of band filling and range from 
metallic to insulating. The key input in the calculations is the 
layer dependent local density of states which we obtain from a 
tight-binding approach for semi-infinite cubic systems. As a result of 
the planar character of the perovskite $t_{2g}$ bands near the Fermi 
level, the reduced coordination number of surface atoms gives rise to 
a significant narrowing of the surface density of those subbands which
hybridize preferentially in planes normal to the surface.
Although the total band width coincides with the one in the bulk,
the effective band narrowing at the surface leads to stronger 
correlation features in the quasi-particle spectra. In particular,
the weight of the quasi-particle peak near $E_F$ is reduced and
the amplitude of the lower and upper Hubbard bands is enhanced,
in agreement with experiments.          
\end{abstract}
\maketitle

\section{introduction}

Strongly correlated materials are presently an active area of
experimental and theoretical investigation\ \cite{imada,georges}.
Angle-resolved photoemission spectroscopy
in principle provides a complete map of the energy and momentum 
dependent quasi-particle states and has therefore been used to
characterize the electronic properties of many fascinating systems.
As a consequence of the finite escape depth of the emitted electron,
however, photoemission spectra inevitably contain information on 
the electronic structure in the bulk and near the surface. 
Both the single- and many-particle features may depend on the
distance from the surface. For example, even in the absence of any
surface reconstruction the reduced coordination of surface atoms
causes a characteristic oscillatory variation of the local density
of states as a function of the layer index, with an appreciable
effective band narrowing right at the surface\ \cite{kalkstein}.
In strongly correlated 
materials it is precisely the width and shape of this local density
of states which, for a given set of on-site Coulomb and exchange 
energies, determine the details of the quasi-particle spectrum.
Accordingly, the relative weights of the quasi-particle peak near 
the Fermi level and of the Hubbard bands may vary with distance 
from the surface. Additional effects can arise due to reconstruction
of the lattice at the surface, due to more pronounced rotations
or distortions of some of the atomic groups within the unit cell,
surface phonons, and due to enhanced electron-electron interaction
caused by less efficient surface screening processes.    

Surface effects in photoemission from transition metal oxides have 
been observed in several systems. Fujioka {\it et al.} \cite{fujioka} 
studied SrRuO$_3$ and noticed characteristic spectral variations caused 
by the frequency dependent mean free path of the photoelectron.
A similar trend was found by Maiti {\it et al.}\ \cite{maiti00} for
the series Ca$_x$La$_{1-x}$VO$_3$. The latter data suggested the highly
interesting situation of a metallic bulk coexisting with an insulating
surface layer. Recently, Maiti {\it et al.}\
\cite{maiti} and Sekiyama {\it et al.}\ \cite{sekiyama} performed 
photoemission measurements on Ca$_x$Sr$_{1-x}$VO$_3$ using a wide 
range of photon energies. These data also reveal striking effects 
associated with the varying amounts of bulk and surface contributions 
to the spectra. Typically, the valence bands in these systems consist 
of a coherent peak near $E_F$ derived from the partially filled 
transition metal $t_{2g}$ bands, and a satellite feature corresponding 
to the lower Hubbard band. Inverse photoemission spectra reveal an 
analogous Hubbard band above $E_F$. In the measurements cited above, 
the weight of the coherent peak diminishes for shorter escape depth, 
while the satellite features (the so-called incoherent peaks) become 
more intense. Photoemission spectra exhibiting a relatively larger 
surface contribution therefore are more strongly correlated 
than bulk spectra. 

This explains the puzzling behavior seen in early photoemission work on 
Ca$_x$Sr$_{1-x}$VO$_3$\ \cite{aiura,inoue95,morikawa} in which Ca doping 
caused a significant suppression of intensity near $E_F$ and therefore 
appeared to drive the system close to a Mott transition. These results
were at odds with the metallic behavior found in various thermodynamic 
measurements independently of Ca concentration\ \cite{onada}. 
Separating bulk and surface contributions by using different photon
energies, Maiti {\it et al.}\ \cite{maiti} and Sekiyama {\it et al.}\ 
\cite{sekiyama} demonstrated that the bulk emission from SrVO$_3$ and
CaVO$_3$ is quite similar, in agreement with the low-frequency bulk 
probes and recent theoretical work\ \cite{nekrasov}. 
The surface spectra of both materials, however, are considerably more 
correlated.

A related example is Sr$_2$RuO$_4$ for which previous photoemission 
spectra seemed to contradict bulk de Haas-van Alphen measurements\ 
\cite{SrRuO1}. Recent experimental and theoretical work proved, however, 
that this discrepancy can be resolved by taking into account the 
lattice reconstruction at the surface of Sr$_2$RuO$_4$ which leads to 
significant changes in the photoemission spectra\,\cite{SrRuO2}. 

To derive reliable information on bulk properties of strongly correlated 
systems using photoemission it clearly is desirable to identify single- 
and many-electron effects associated with the surface. 
In the present work we study these effects for three perovskite 
materials with widely different band fillings: the metallic compounds
Ca$_{x}$Sr$_{1-x}$VO$_3$ ($d^1$) and  SrRuO$_3$ ($d^4$), and the series 
Ca$_{x}$La$_{1-x}$VO$_3$ which is insulating for $x=0$ ($d^2$), 
but metallic for $x=0.5$ ($d^{1.5}$).
We evaluate the quasi-particle self-energy using the dynamical mean 
field theory based on the multi-orbital Monte Carlo method  
\ \cite{georges,vollhardt,pruschke}. The important input in these 
many-body calculations is the layer dependent local density of states 
which we derive from a tight-binding scheme for semi-infinite systems\
\cite{kalkstein}. We show that the surface leads to an effective 
narrowing of the density of states of those bands hybridizing mainly 
in atomic planes normal to the surface. As a result, correlation 
effects at the surface are more pronounced than in the bulk. Such a
trend had first been predicted by Potthoff and Nolting\ \cite{potthoff}
who studied the metal-insulator phase diagram for a semi-infinite 
simple cubic $s$ band at half filling. Here we calculate the self-energy  
for multi-band systems using realistic local densities of states for 
several cubic perovskite materials and find qualitative agreement with 
photoemission data\ \cite{fujioka,maiti00,maiti,sekiyama}. Preliminary
results on SrVO$_3$ were published earlier\ \cite{lie03}.  

This paper is organized as follows. In Section II we focus on the 
single-particle electronic properties of SrVO$_3$ which is can be 
taken as representative of perovskite materials. In particular, 
we discuss the evaluation of the layer dependent local density of states 
for semi-infinite SrVO$_3$. Section III provides the main elements of the
calculation of the multi-orbital self-energy in the bulk and at the 
surface. The quasi-particle spectra of Ca$_x$Sr$_{1-x}$VO$_3$,
SrRuO$_3$, and Ca$_x$La$_{1-x}$VO$_3$ are presented in Section IV. 
Section V contains the summary.

\section{Electronic Structure: \ SrVO$_3$ }

In this section we discuss the bulk and surface electronic 
properties of SrVO$_3$. This system can be considered as a prototype
of a cubic perovskite material since its one-electron structure is
relatively simple, with one $d$ electron per transition metal ion.
The other systems can be qualitatively understood in terms of
these properties by accounting for different occupations: $d^4$ for
SrRuO$_3$, and $d^{2-x}$ for Ca$_x$La$_{1-x}$VO$_3$. 

Self-consistent electronic structure calculations for bulk SrVO$_3$ 
within the local density approximation (LDA)\, \cite{takegahara} 
show that the conduction bands near the Fermi level consist of three 
degenerate $t_{2g}$ bands derived from V$^{4+}$ ($3d^1$) ions. 
The filled O 2p bands are separated from the $t_{2g}$ levels by a gap
of about 1~eV, and the cubic crystal field of the V-O octahedron 
shifts the V $e_g$ bands above the $t_{2g}$ bands. Because of the 
cubic symmetry, the $t_{2g}$ bands can be represented via a 
tight-binding Hamiltonian with diagonal elements
\begin{eqnarray}
 h_{xy,xy}(k)&=& e_d + t_0 (c_x + c_y) + t_1 c_x c_y  \nonumber\\
             && + \ [t_2  + t_3(c_x + c_y) + t_4 c_x c_y] c_z\,,       
\end{eqnarray}
where $c_i = 2\cos(k_ia),\, i=x,y,z$ and $a$ is the lattice constant. 
Cyclic permutations yield $h_{xz,xz}(k)$ and $h_{yz,yz}(k)$.
The $t_i$ denote effective hopping integrals representing the V-O-V 
hybridization, where $t_{0,2}$, $t_{1,3}$, and $t_4$ specify the 
interaction between first, second and third neighbors, respectively.
For symmetry reasons off-diagonal elements arise only between second 
and third nearest neighbors and are of the form \,$h_{xy,xz}(k) = 
- t'_1 s_y s_z - t'_2 c_x s_y s_z$, where $s_i = 2\sin(k_ia)$, i.e.,   
they vanish at the high-symmetry points.
Since the coefficients $t'_{1,2}$ are very small we neglect these 
off-diagonal elements so that the energy bands are 
given by \,$\epsilon_i(k) =  h_{i,i}(k)$, with \,$i=xy,xz,yz$.
The tight-binding parameters $e_d$ and $t_0\cdots t_4$ can easily 
be found by fitting the LDA energies at high-symmetry points of the 
bulk Brillouin Zone. 
 
\begin{figure}[t]
  \begin{center}
   \includegraphics[width=5cm,height=8cm,angle=-90]{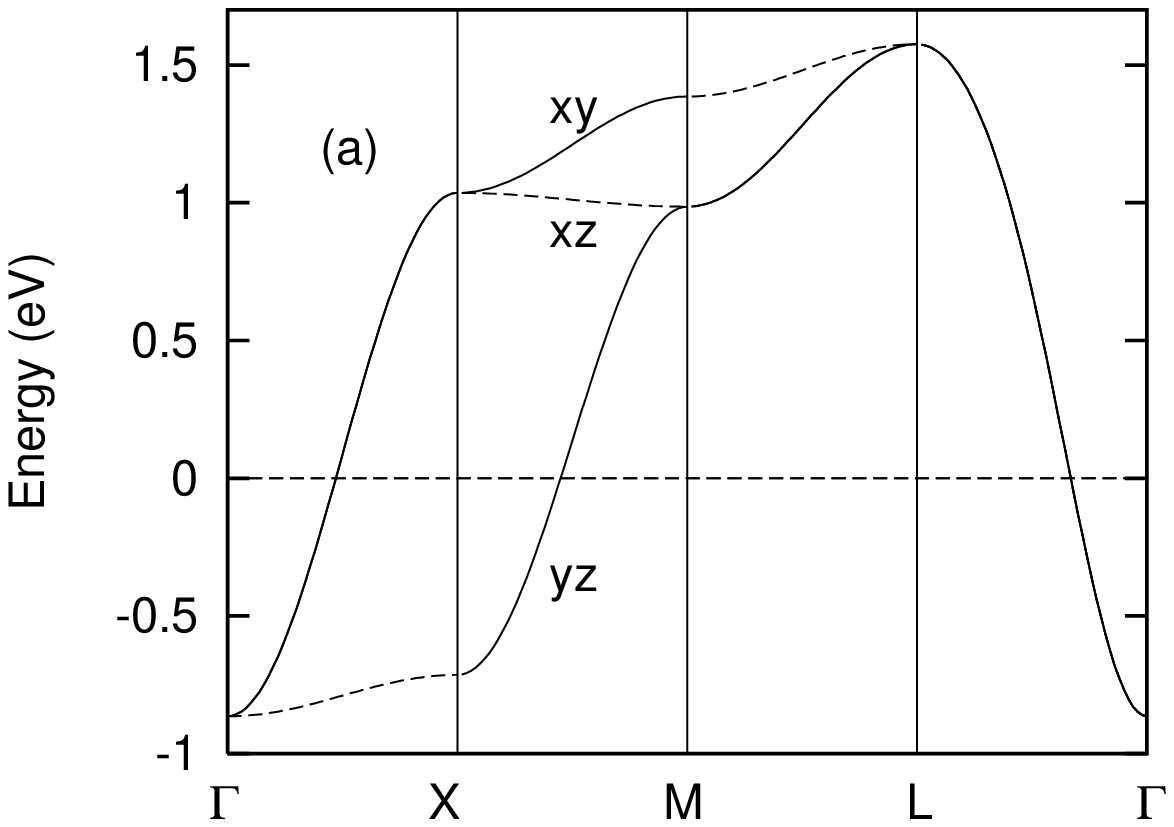}
   \includegraphics[width=5cm,height=8cm,angle=-90]{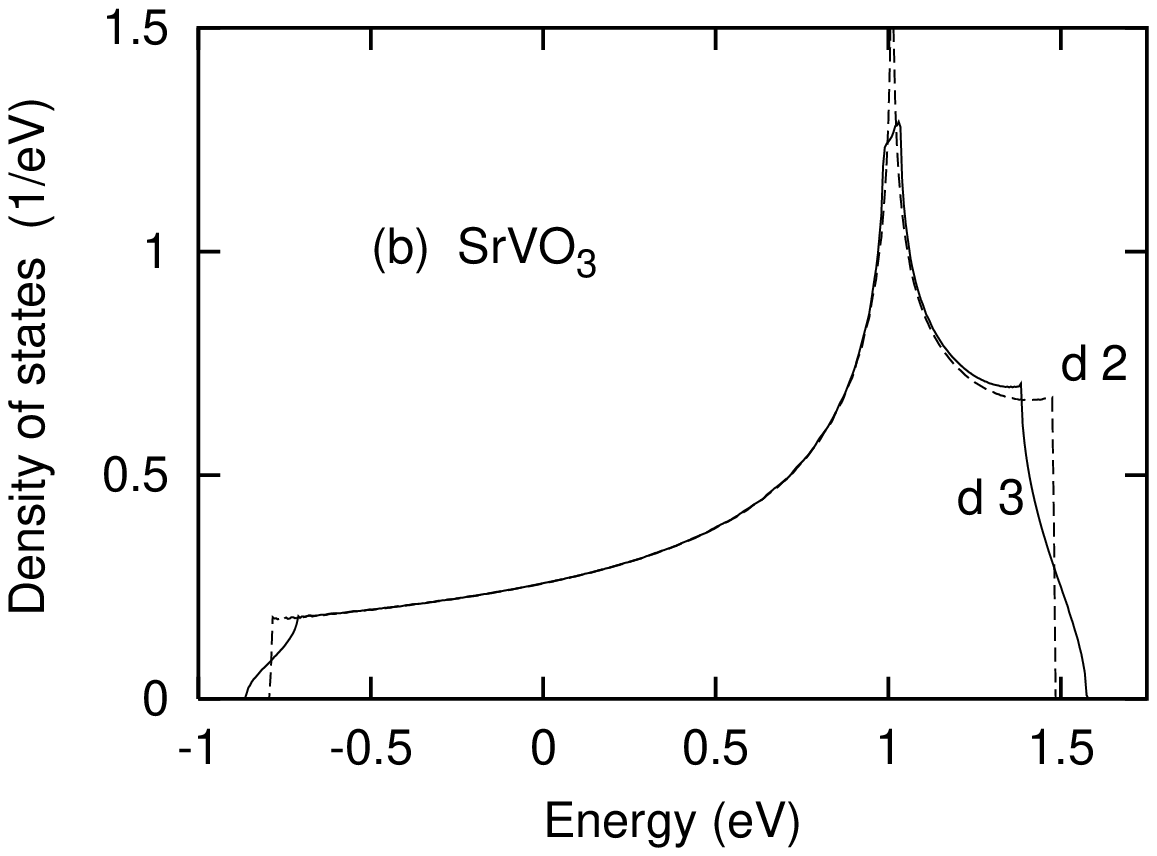}
  \end{center}
\caption{
(a) Tight-binding fit to LDA $t_{2g}$ bulk bands of SrVO$_3$ ($3d^1$).
Solid lines: dispersion within the orbital plane; dashed lines: 
dispersion perpendicular to the orbital plane  ($E_F=0$). 
(b) Solid curve: density of states of SrVO$_3$  $t_{2g}$ bulk bands.
Dashed curve: analogous two-dimensional density of states
obtained by neglecting inter-planar hopping integrals.    
}\end{figure}

Fig.~1\,(a) shows the $t_{2g}$ bulk bands of SrVO$_3$ along the main 
symmetry directions. The conduction bands in the cubic perovskite 
structure consist of three nearly non-hybridizing $t_{2g}$ bands. 
Each of these bands is approximately two-dimensional representing weakly 
coupled atomic planes. Typically, the inter-planar band width is about 
20 times smaller than the intra-planar band width. To indicate the
pronounced planar character of these bands we denote via solid lines 
the strong dispersion within the plane spanned by an orbital and by 
dashed lines the much weaker out-of-plane dispersion. According to this 
nearly two-dimensional electronic structure, the bulk density of states 
$\rho_b(\omega)$ exhibits the characteristic main peak related to the 
van Hove singularity at the X point of the Brillouin Zone. 
This is shown in Fig.~1\,(b) where the bulk density 
of states is compared with the two-dimensional density obtained by 
setting the inter-planar hopping integrals $t_{2,3,4}$ equal to zero. 
The asymmetric shape of both distributions follows from the 
second-neighbor hopping terms $\sim t_1$. 
The overall shape of the bulk density agrees well with the one obtained 
from LAPW calculations\,\cite{takegahara}.

Note that as a result of the $3d^1$ configuration, the Fermi surface of 
SrVO$_3$ consists of three nearly perfect intersecting cylinders containing 
the $d_{xy}$, $d_{xz}$ and $d_{yz}$ states. This peculiar shape was recently 
observed also in de Haas--van Alphen measurements of CaVO$_3$\ 
\cite{inoue02} in spite of pronounced orthorhombic distortions.
 
In the cubic environment, the three $t_{2g}$ bands have identical density 
of states. At the surface, the bulk degeneracy is 
lifted since only the $d_{xy}$ band exhibits strong dispersion within the 
plane of the surface (the $z$ direction specifies the surface normal) whereas 
the $d_{xz}$ and $d_{yz}$ bands disperse primarily within atomic planes
perpendicular to the surface plane. Thus, the local density of states of 
the $d_{xy}$ band in the first layer is similar to the bulk 
density, while that of the $d_{xz,yz}$ bands is modified by the 
reduced coordination number in the $z$ direction. To evaluate these 
surface densities we use a Green's function formalism\ \cite{kalkstein} 
for semi-infinite tight-binding systems.

Neglecting again the weak hybridization between $t_{2g}$  states,
the local density of states of band $i\equiv xy,xz,yz$ is given by 
\begin{equation}
 \rho_{i,n}(\omega) = \frac{1}{\pi}\,\sum_{k_\parallel}\ {\rm Im}
                  \,G_{i,n}(k_\parallel,\omega)
\end{equation}
where $n\ge1$ denotes the layer index and the Green's function 
can be conveniently determined from the expression 
\begin{equation}
G_{i,n}(k_\parallel,\omega) = \frac{i}{\mu}\left[1+ 
             \left(\frac{i\mu+\Omega}{2T_i}\right)^m\!\!
             \left(\frac{i\mu+\Omega-\Delta}{i\mu-\Omega+\Delta}\right)
                                         \right]
\end{equation}
where $m=2n-2$, \,$\mu = (4T_i^2 - \Omega)^{1/2}$\, and
\,$\Omega = \omega - W_i$. The parameter $\Delta$ denotes a surface 
potential.
$W_i$ and $T_i$ represent intra- and inter-planar contributions 
to the $t_{2g}$ band energies:
\begin{eqnarray}
   W_{xy} &=&  e_d + t_0 (c_x + c_y) + t_1 c_x c_y  \nonumber\\
   W_{xz} &=&  e_d + t_0  c_x + t_2 c_y + t_3 c_x c_y  \\
   W_{yz} &=&  e_d + t_0  c_y + t_2 c_x + t_3 c_x c_y  \nonumber
\end{eqnarray}
and
\begin{eqnarray}
   T_{xy} &=&  t_2 + t_3(c_x + c_y) + t_4 c_x c_y    \nonumber\\ 
   T_{xz} &=&  t_0 + t_1 c_x + t_3 c_y + t_4 c_x c_y  \\
   T_{yz} &=&  t_0 + t_1 c_y + t_3 c_x + t_4 c_x c_y \,.\nonumber 
\end{eqnarray}
How SrVO$_3$ is terminated at the surface, especially the charge 
state of the V ions in the first layer, is not yet known. Thus, we 
assume all tight-binding parameters to coincide with those in the bulk. 
The surface potential is chosen to ensure charge neutrality (see below). 

\begin{figure}
\begin{center}     
   \includegraphics[height=8cm,width=5.0cm,angle=-90]{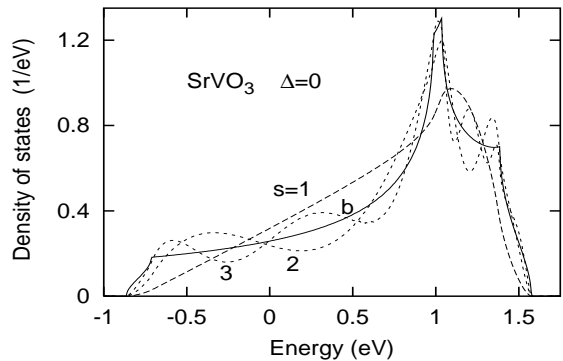}
\end{center}
\caption{
Layer dependent local density of states $\rho_{i,n}(\omega)$ of 
out-of-plane $d_{xz,yz}$ bands for first three atomic planes of 
SrVO$_3$ ($E_F=0$). The surface potential $\Delta$ is set equal to zero.  
Solid curve: isotropic bulk density of states 
$\rho_b(\omega)$. The local density of the intra-planar $d_{xy}$ states 
(not shown) is similar to the bulk density, even in the surface layer.   
}\end{figure}

Fig.~2 compares the layer dependent local density of states 
$\rho_{i,n}(\omega)$ of the $d_{xz,yz}$ bands with the isotropic bulk 
density $\rho_b(\omega)$. The $d_{xz,yz}$ density in the first plane 
is seen to be less sharply peaked and more narrow than $\rho_b(\omega)$ 
although their total widths are identical. In particular, 
$\rho_{xz,1}(\omega)$ rises almost linearly below $E_F$ in contrast 
to the plateau-like shape of $\rho_b(\omega)$. Thus, the surface spectral 
weight is reduced at low and high frequencies and enhanced at intermediate 
frequencies. The local density
of $d_{xz,yz}$ states in the deeper lying layers approaches the bulk 
density rather quickly, the main effect consisting in an oscillatory
distortion of the spectral shape rather than any appreciable band
narrowing. The rapid convergence towards the bulk density is to be 
expected because of the tight-binding character of the $t_{2g}$
bands resulting from the short range of the hopping integrals. 
The local density of the in-plane $d_{xy}$ states depends only 
weakly on the layer index and nearly coincides with the bulk density
even in the first layer (see below). 

It would of course be desirable to perform
self-consistent electronic structure calculations for semi-infinite
SrVO$_3$ since they should provide a more accurate description of the
density of states in the surface region. Nevertheless, we believe that
the key effect discussed here within the simplified tight-binding 
approach, namely, the preferential band narrowing of the $d_{xz,yz}$ 
states, will hold quite generally.             

The single particle properties of the remaining perovskite systems 
discussed in this work are closely related to those of SrVO$_3$.
They differ essentially by the degree of filling of the $t_{2g}$ bands,
and by the deviation from cubic symmetry via orthorhombic distortions
of the oxygen octahedra surrounding the transition metal ions. These
distortions primarily broaden the van Hove singularity of the density
of states but have only a minor effect on the overall width of the 
$t_{2g}$ bands\ \cite{mazin,nekrasov}. 
We discuss these differences in more detail in Section IV.
 
\section{quasi-particle spectra}

To analyze the experimental photoemission data we evaluate the 
quasi-particle spectra by taking into account local Coulomb interactions.
According to the semi-infinite one-electron properties discussed above 
we are dealing with a non-isotropic system where two narrow $d_{xz,yz}$
bands interact with a wider  $d_{xy}$ band. 
In the first atomic layer this difference in 
effective band width is most pronounced and it diminishes rapidly towards 
the interior of the system. This situation is reminiscent of the one
in the layer perovskite Sr$_2$RuO$_4$, which essentially consists of 
Ru sheets containing two nearly one-dimensional $d_{xz,yz}$ bands
interacting with a wide intra-planar $d_{xy}$ band. The peculiar
interesting feature of the latter system is the fact the on-site
Coulomb energy lies between the single-particle widths of the non-%
degenerate $t_{2g}$ bands: \,$W_{xz,yz} < U < W_{xy}$\ \cite{liebsch}. 
In the present case, on the other hand, the difference between the 
$d_{xz,yz}$ and $d_{xy}$ bands at the surface is less pronounced so 
that $W_i<U$ for all three bands. Nevertheless, since in the first 
layer the effective width of $d_{xz,yz}$ states is reduced, the effect 
of Coulomb correlations on the surface bands should be stronger than 
on the wider bulk bands.   
 
The key quantity characterizing the effect of Coulomb correlations 
on quasi-particle spectra is the self-energy which we evaluate here
using the dynamical mean field theory (DMFT)
\ \cite{georges,vollhardt,pruschke}. 
A full description of correlations near the surface 
would be exceedingly complicated since, in principle, it would require
a mixed momentum/real space approach in order to handle the loss of
translational symmetry normal to the surface. This could be accomplished
using a cluster formalism in which the semi-infinite system is represented
via a slab of finite thickness. Unfortunately, the planar character of
the $t_{2g}$ states leads to a slowly convergent local density of 
$d_{xz,yz}$ states, with many spurious $1/\sqrt{\omega}$ van Hove 
singularities stemming from the quasi-one-dimensional hopping along atomic 
rows parallel to the surface. On the other hand, a cluster generalization 
of the DMFT is feasible today only for very small cluster size.

To achieve a qualitative understanding of correlations at SrVO$_3$ 
surfaces we ignore the momentum variation of the self-energy and 
assume that, for a given layer, it depends only on the local density
of states within that layer\ \cite{potthoff}. 
Since we neglect the weak hybridization between $t_{2g}$ orbitals, the 
self-energy is diagonal in orbital space. 
To evaluate the self-energy elements $\Sigma_i(\omega)$ we use the 
dynamical mean field approach, in which $\Sigma_i$ is a functional of
the bath Green's function \,${\cal G}_i^{-1} = G_i^{-1} + \Sigma_i$,
where the local  $G_i$ is given by   
\begin{equation}
  G_i(i\omega_n) = \int^{\infty}_{-\infty}\! d\omega\
      \frac{\rho_i(\omega)}{i\omega_n+\mu-\Sigma_i(i\omega_n)-\omega}. 
\end{equation}
The Matsubara frequencies are denoted by $\omega_n$ and $\mu$ is the 
chemical potential. 
On-site Coulomb correlations are treated using the self-consistent multiband 
Quantum Monte Carlo (QMC) method (for a review, see Ref.\ \cite{georges}).
The temperature of the simulation was 125~meV ($\beta=8$). Several runs
using 64 imaginary time slices and $10^5$ Monte Carlo sweeps were carried 
out. The quasi-particle density of states 
\,$N_i(\omega)=-{\rm Im}\,G_i(\omega)/\pi$ 
was obtained via maximum entropy reconstruction\ \cite{jarrell}.

In principle, we are now faced with a set of coupled impurity problems, 
where the layer-dependent baths determine the self-energies of each
atomic plane. This problem could be solved iteratively until 
self-consistency is achieved and a common chemical potential is 
found for the bulk and at the surface. Unfortunately, for multi-band
systems this iteration procedure would be computationally extremely 
demanding and will be neglected here.
On the other hand, previous work\ \cite{potthoff} showed that the 
self-energy is mainly governed by local Coulomb correlations 
and that further changes due to interlayer coupling are rather small. 
Our assumption of charge neutrality within the surface layer, and the
fact that correlations in the bulk and at the surface do not differ 
substantially, most likely also reduce the importance of interlayer 
effects. The key layer dependent input in our QMC-DMFT calculation
is therefore the one-electron local density of states. For a given set 
of local Coulomb and exchange energies the width and shape of this local 
density then determines the details of the quasi-particle spectrum.  

\begin{figure}[htbp]
  \begin{center}
   \includegraphics[width=5cm,height=8cm,angle=-90]{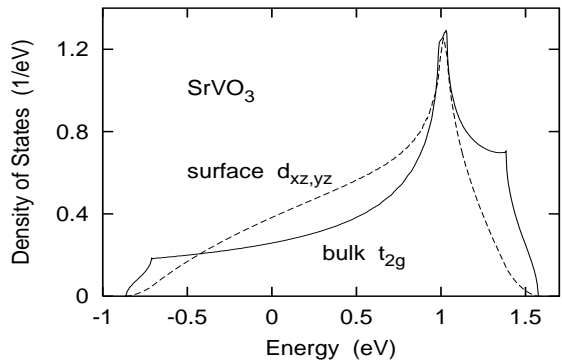}
  \end{center}
\caption{
Local density of $d_{xz,yz}$ states of first layer of SrVO$_3$ ($d^1$) 
in the presence of a weak surface potential $\Delta$ to ensure charge 
neutrality (dashed curve). The $t_{2g}$ bulk density (solid curve) is 
shown for comparison. The density of the in-plane $d_{xy}$ states in 
the first layer (dotted curve) is nearly identical to the bulk density. 
}\end{figure}

\section{Results and Discussion}

\subsection{SrVO$_3$}

The layer dependent local density of states of SrVO$_3$ shown in Fig.~1(b)
is calculated in the absence of any surface potential. Thus, the occupation 
numbers also vary with layer index. However, since local Coulomb correlations
are very sensitive to the degree of band filling, it seems appropriate to 
enforce charge neutrality by adjusting $\Delta$ accordingly. Fig.~3 shows 
the resulting local density of states for the surface layer. Although the
main peak now coincides more with the bulk van Hove singularity, the  
effective narrowing is similar to the one in Fig.~1(b). Note that the 
surface potential is sufficiently weak so that no surface states are split 
off below the band. Thus, the total band width is the same as in the bulk.   
No surface potential is needed for the $d_{xy}$ states since their local
density is practically identical to the bulk $t_{2g}$ density.

\begin{figure}[t!]
\begin{center}     
   \includegraphics[height=7cm,width=4.0cm,angle=-90]{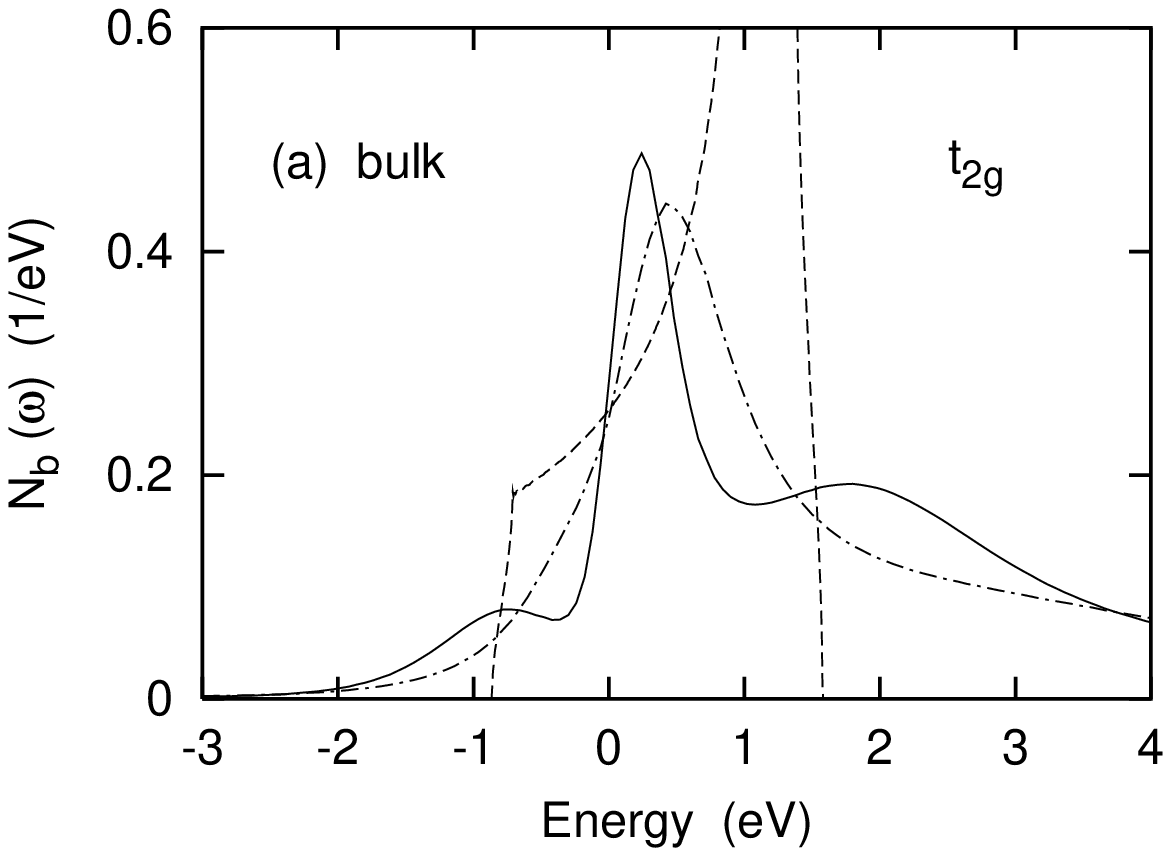}
   \includegraphics[height=7cm,width=4.0cm,angle=-90]{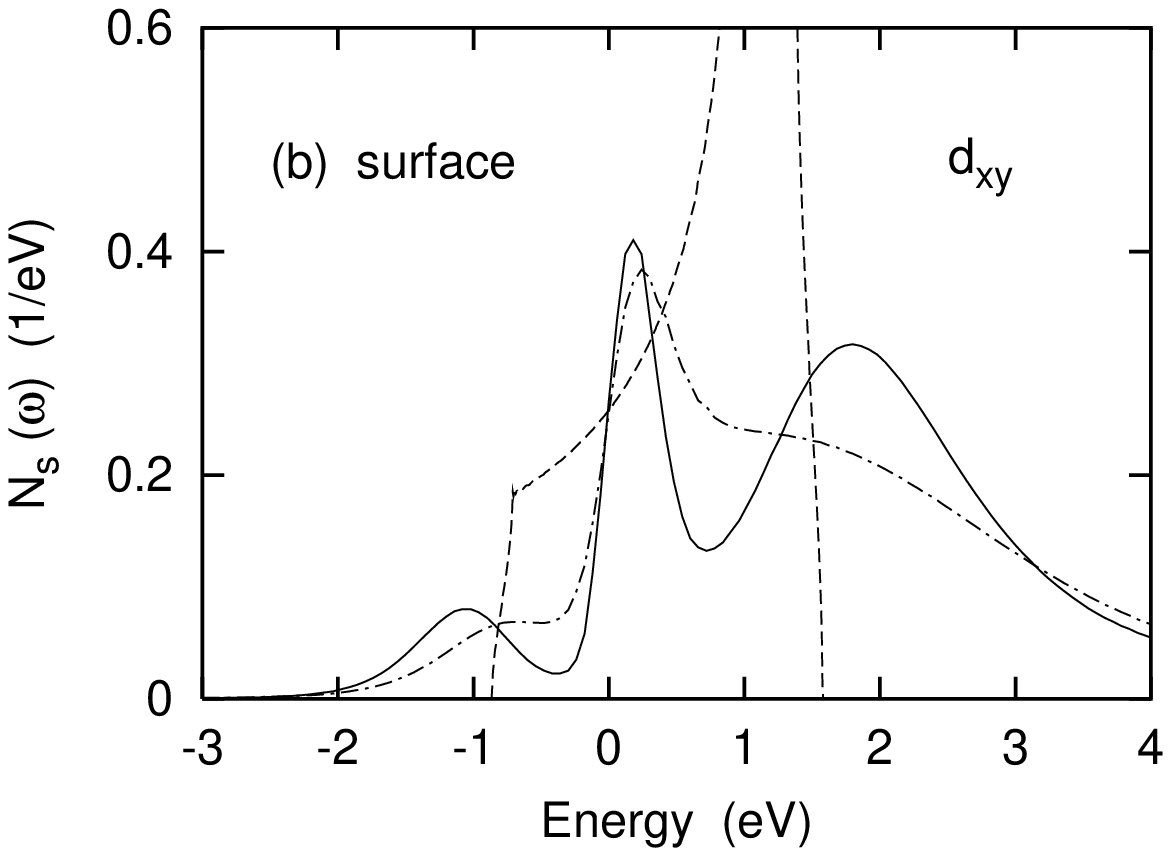}
   \includegraphics[height=7cm,width=4.0cm,angle=-90]{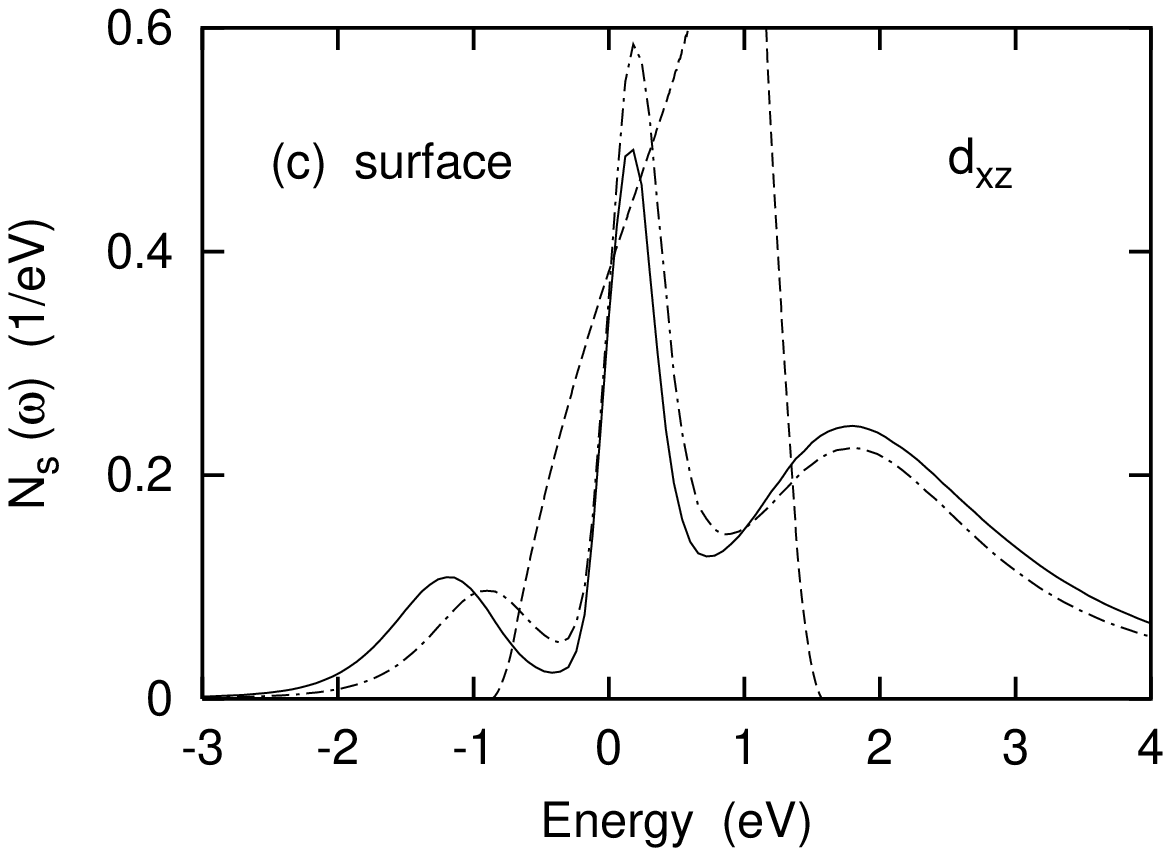}
   \includegraphics[height=7cm,width=4.0cm,angle=-90]{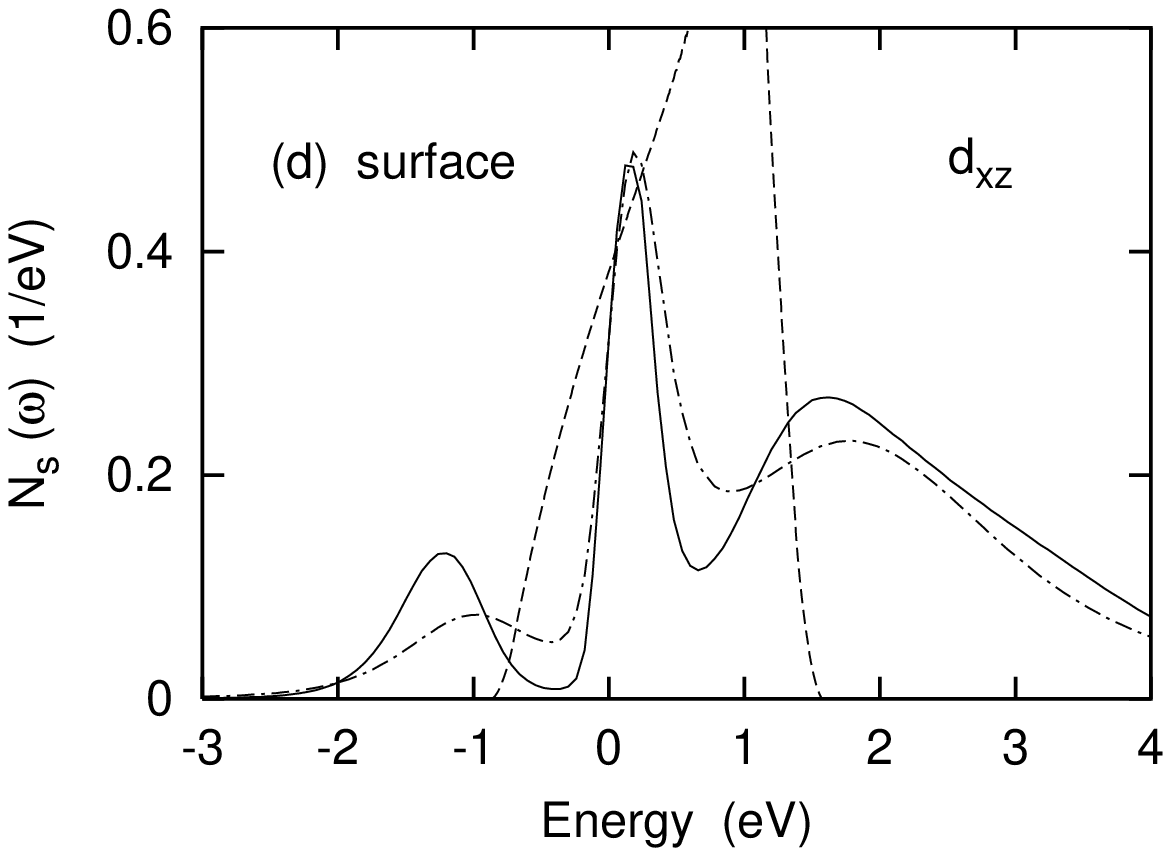}
\end{center}
\caption{
Quasi-particle density of states $N_i(\omega)$ of SrVO$_3$ ($d^1$) 
derived from DMFT.
(a) bulk $t_{2g}$ states; (b) surface $d_{xy}$ states; (c) surface 
$d_{xz,yz}$ states; (d) fictitious isotropic surface $d_{xz,yz}$ states 
(see text).
Solid curves: $U=4.3$\,eV, dot-dashed curves:  $U=4.0$\,eV ($J=0.7$\,eV).
Dashed curve: bare densities of states (see Fig.~3).
}\end{figure}

Fig.~4\,(a) shows the bulk quasi-particle density of states of SrVO$_3$  
for two Coulomb energies in the region where the Hubbard bands which are 
seen as satellites in photoemission begin to emerge: $U=4.0$\,eV and 
$4.3$\,eV. The exchange energy is $J=0.7$\,eV\ \cite{zaanen}. 
These results show that in the bulk $U$ must be larger than 4\,eV
to obtain the satellite observed in photoemission spectra 
\cite{maiti,sekiyama}. The peak near 2~eV above $E_F$ agrees with 
inverse photoemission data\ \cite{morikawa}. Although for $U=4$\,eV 
there is considerable correlation-induced band narrowing and an 
emerging satellite shoulder, the larger $U$ to yields 
an even narrower coherent feature near $E_F$, with the missing weight 
shifted to the lower and upper Hubbard bands. 

Note that $N_b(E_F)=\rho_b(E_F)$ which follows (at $T=0$) from the local 
approximation implicit in the DMFT for isotropic systems\ \cite{pinning}. 
The most recent photoemission data\ \cite{sekiyama} confirm this result. 
The bulk spectra shown in Fig.~4\,(a) qualitatively agree with DMFT-LDA 
results for SrVO$_3$ by Nekrasov {\it et al.}\ \cite{nekrasov} who 
employed somewhat larger values for $U$ and $J$. They are also consistent
with previous spectra for the $3d^1$ perovskite La$_x$Sr$_{1-x}$TiO$_3$\ 
\cite{nekrasov00} which exhibits a similar $t_{2g}$ bulk density of states.  

The surface quasi-particle spectra for SrVO$_3$ are shown in Fig.~4\,(b) 
and (c). The lower Hubbard peak of the $d_{xz,yz}$ states in (c)
is clearly visible already for $U=4$\,eV because of the narrower local 
density of states in the first layer. A larger $U$ shifts the 
satellite to higher binding energies. The comparison with the spectra
shown in Fig.~4\,(a) demonstrates that correlation effects for a fixed 
value of $U$ are stronger at the surface than in the bulk: The coherent 
peak near $E_F$ is narrower at the surface and the incoherent satellite 
feature is more pronounced than in the bulk, in agreement with 
experiment\ \cite{maiti,sekiyama,fujioka}. 

The surface quasi-particle density of $d_{xy}$ states in Fig.~4\,(b) 
is intermediate between $N_b(\omega)$ and $N_s(\omega)$ for 
$d_{xz,yz}$. Although there is little single-electron hybridization 
between $t_{2g}$ bands, the local Coulomb interaction mixes them so that 
the $d_{xy}$ surface spectrum involves contributions arising from the 
more strongly correlated $d_{xz,yz}$ states. Another consequence
of the anisotropic surface self-energy is the 
correlation-induced charge transfer between subbands\ \cite{liebsch}. 
Here, we find that 0.06 electrons are shifted from the $d_{xy}$ to 
the $d_{xz,yz}$ bands. Also, the quasi-particle partial densities of 
states at $E_F$ do not need to coincide with the bare partial densities. 
The coupling between narrow and wide bands is a genuine multi-band effect 
and underlines the fact that single-particle bands in the presence of 
local Coulomb interactions cannot be considered independently. 

For illustrative purposes we show in Fig.~4\,(d) the analogous quasi-particle 
spectra for a hypothetical isotropic system whose density of states is given 
by the surface $\rho_{xz,yz}$. Evidently, correlation effects are slightly 
more pronounced than at the surface of SrVO$_3$ where two such narrow bands
interact with the wider $d_{xy}$ band. All spectra shown Fig.~4 are based
on single-particle densities of identical total width but with progressively
more orbital components exhibiting narrower spectral shape. Accordingly, the
weight of the quasi-particle peak near $E_F$ decreases systematically and
the upper and lower Hubbard bands become successively more intense.         

Note that the quasi-particle spectra in Fig.~4 are highly asymmetric and
that the gap between the central peak and the lower satellite appears at
smaller $U$ than the upper gap. This behavior is a consequence of the 
asymmetric bulk and surface density of states, and of the low band
filling in this system (1/6 per spin band). Thus, despite the typical
overall 3-peak structure which is characteristic of many highly correlated 
quasi-particle distributions, their is a large degree of spectral detail 
which is characteristic of this particular material.

We also point out that the many-body reduction of the quasi-particle band 
width is very much larger than the surface-induced one-electron band 
narrowing. 
On the other hand, since the on-site Coulomb energy is not far from the 
critical value for a metal-insulator transition, the band narrowing 
substantially enhances the influence of correlations at the surface.   

It would be interesting to perform angle-resolved photoemission 
measurements to determine the correlation-induced band narrowing  
of the SrVO$_3$ $t_{2g}$ bands. For instance the energy at $\bar\Gamma$ 
should be only a few tenths of an eV below $E_F$ instead of 1\,eV 
as predicted by the LDA. Accordingly, the true $t_{2g}$ bands should be 
considerably flatter than the single-particle bands. Also, measurements 
using polarized light could help to separate correlation effects in the 
$d_{xy}$ and $d_{xz,yz}$ bands.
 
\subsection{CaVO$_3$}

In striking contrast to thermodynamic measurements earlier photoemission 
data on Ca$_x$Sr$_{1-x}$VO$_3$\ \cite{aiura,inoue95,morikawa}
indicated considerably more pronounced correlations in CaVO$_3$ 
than in SrVO$_3$. These observations even led to speculations as to
whether CaVO$_3$ might be close to at Mott transition.  
However, recent photoemission measurements taken over
a wide range of photon energies \cite{maiti,sekiyama} demonstrated 
that this discrepancy can be resolved by carefully separating bulk 
and surface contributions to the spectra by using widely different
photon energies. According to these results,
the bulk emission from CaVO$_3$ is quite similar to that from SrVO$_3$,
the CaVO$_3$ spectra being only slightly more correlated. Thus, the
new data are consistent with low-frequency bulk measurements. 

While SrVO$_3$ is a cubic perovskite with a V-O-V bond angle of $180^o$, 
the oxygen octahedra in CaVO$_3$ are distorted, so that the V-O-V bond 
angle is reduced to $162^o$, implying a weaker indirect hopping between 
$d$ orbitals and a reduction of the  $d$ electron band width. As the most 
recent LDA calculations for CaVO$_3$ show\ \cite{nekrasov}, this decrease 
in total band width amounts to only about 4\,\% (from 2.6 to 2.5~eV).
A more striking consequence of the distortion is the considerable broadening
of the van Hove singularity and the skewing of its spectral weight to lower 
frequencies. Similar results had been obtained in LDA calculations for 
SrRuO$_3$ and CaRuO$_3$\ \cite{mazin}, where orthorhombic distortions also 
cause a  
narrowing of the $t_{2g}$ bands and a broadening of the van Hove singularity.
As the LDA-DMFT bulk calculations for CaVO$_3$ by Nekrasov {\it et al.}\ 
\cite{nekrasov} show, these changes of the density of states lead to 
a small reduction of the quasi-particle peak near $E_F$ and to slightly 
more intense Hubbard bands compared to those in SrVO$_3$, in agreement
with the data.    

Considering the enhancement of Coulomb correlations at the SrVO$_3$ surface
discussed in the previous section, and the fact that in the bulk CaVO$_3$
is already somewhat more correlated than SrVO$_3$, it is now plausible that 
correlation effects should be even stronger at the CaVO$_3$ surface, as 
indeed observed. Since the local Coulomb energy is not far from the critical
value for a metal-insulator transition, the tendency for enhanced surface
correlations in  CaVO$_3$ should in fact be stronger than for  SrVO$_3$.
This effect is entirely a consequence of the one-electron band 
narrowing of the $d_{xz,yz}$ subbands. Additional effects could be caused 
by surface reconstruction or by more pronounced tilting of oxygen octahedra 
at the CaVO$_3$ surface which would lead to weaker in-plane hopping 
for the $d_{xy}$ bands.  For a more detailed analysis it would be 
necessary to carry out electronic structure calculations for the CaVO$_3$ 
surface in order to study orthorhombic distortions and possible 
reconstructions. In addition, theoretical estimates of a possible increase 
of $U$ at the surface due to reduced screening would be useful. 

\subsection{SrRuO$_3$}

The ruthenates with perovskite-based crystal structure have attracted 
considerable attention during the past years because of a variety of 
fascinating properties. SrRuO$_3$ is the only known ferromagnetic 
metal among the $4d$ oxides, with a Curie temperature of 160 K. 
The one-electron density of states is similar to that of SrVO$_3$ 
except for the $d^4$ occupancy of the Ru derived $t_{2g}$ bands\ 
\cite{mazin}. In a perfect cubic system the Fermi level 
would nearly coincide with the van Hove singularity and the flat 
bands in this energy region. However, the oxygen octahedra in SrRuO$_3$ 
are tilted by about $8^o$ which leads to a substantial broadening of
the van Hove singularity. Nevertheless, the density of states near
$E_F$ remains large, giving rise to a large Stoner parameter. 
Doping with Ca causes further tilting of the oxygen octahedra.
As a result the density of states at $E_F$ is reduced even further   
and the magnetism is easily suppressed. The layered ruthenate 
Sr$_2$RuO$_4$ is a paramagnetic metal and is currently intensively
studied because of its unconventional p-wave superconductivity
below 1.5~K\ \cite{phystoday}.

\begin{figure}[t!]
  \begin{center}
  \includegraphics[width=5.0cm,height=8cm,angle=-90]{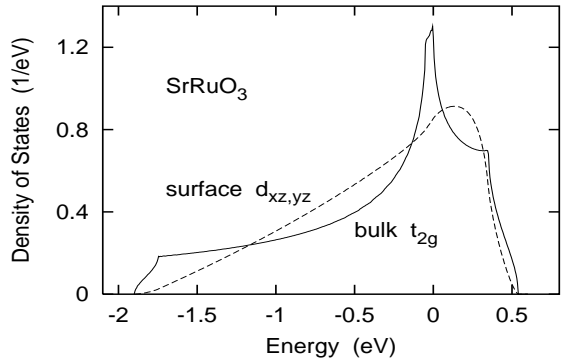}
  \end{center}
\caption{
Solid curve: isotropic bulk density of states of cubic SrRuO$_3$;
dashed curve: local density of  $d_{xz,yz}$ states in first layer  
($E_F=0$).
}\end{figure}

The unusual magnetic properties of the ruthenates have stimulated
wide-ranging investigations of the correlation effects in these 
materials. Transport measurements\ \cite{kostic} as well as theoretical 
studies\ \cite{laad} on SrRuO$_3$ seemed to suggest non-Fermi-liquid 
behavior. Nevertheless, recent transport data on highly ordered 
samples\ \cite{capogna} are consistent with Fermi-liquid theory, but 
demonstrate a remarkable sensitivity to thermal and impurity scattering.
 
Photoemission spectra on SrRuO$_3$\ \cite{fujioka} revealed marked
spectral changes for different photon energies. At 1254~eV emission
from the Ru $4d$ bands exhibits pronounced weight at $E_F$ and a
maximum at about 1.5~eV binding energy. At 100~eV, on the other hand,
the spectral weight at $E_F$ is much smaller, and the intensity then
grows down to 2~eV binding energy. (The region farther below is 
dominated by O $2p$ emission.) Thus, the relatively more important 
surface contribution caused by the shorter escape depth of the 
photoelectron makes the spectrum appear more correlated than at
energies governed by bulk emission.

\begin{figure}[t!]
\begin{center}     
   \includegraphics[height=7cm,width=4.0cm,angle=-90]{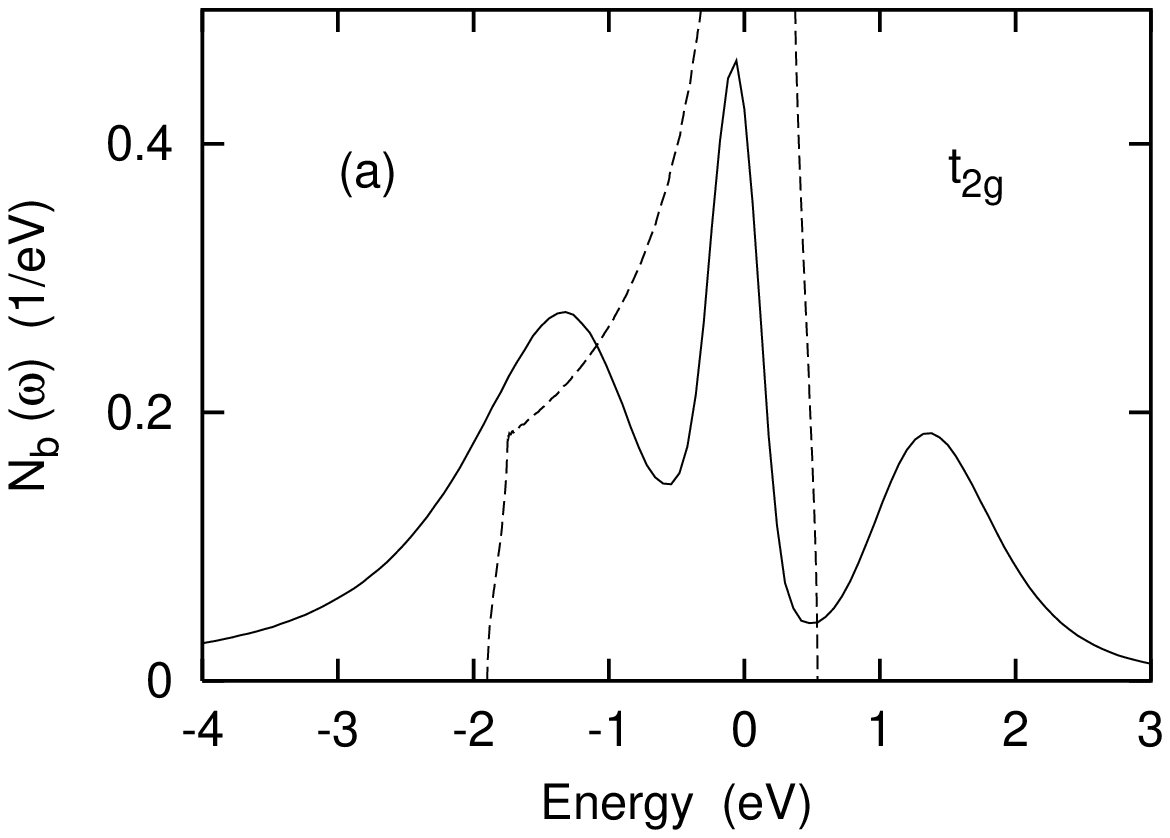}
   \includegraphics[height=7cm,width=4.0cm,angle=-90]{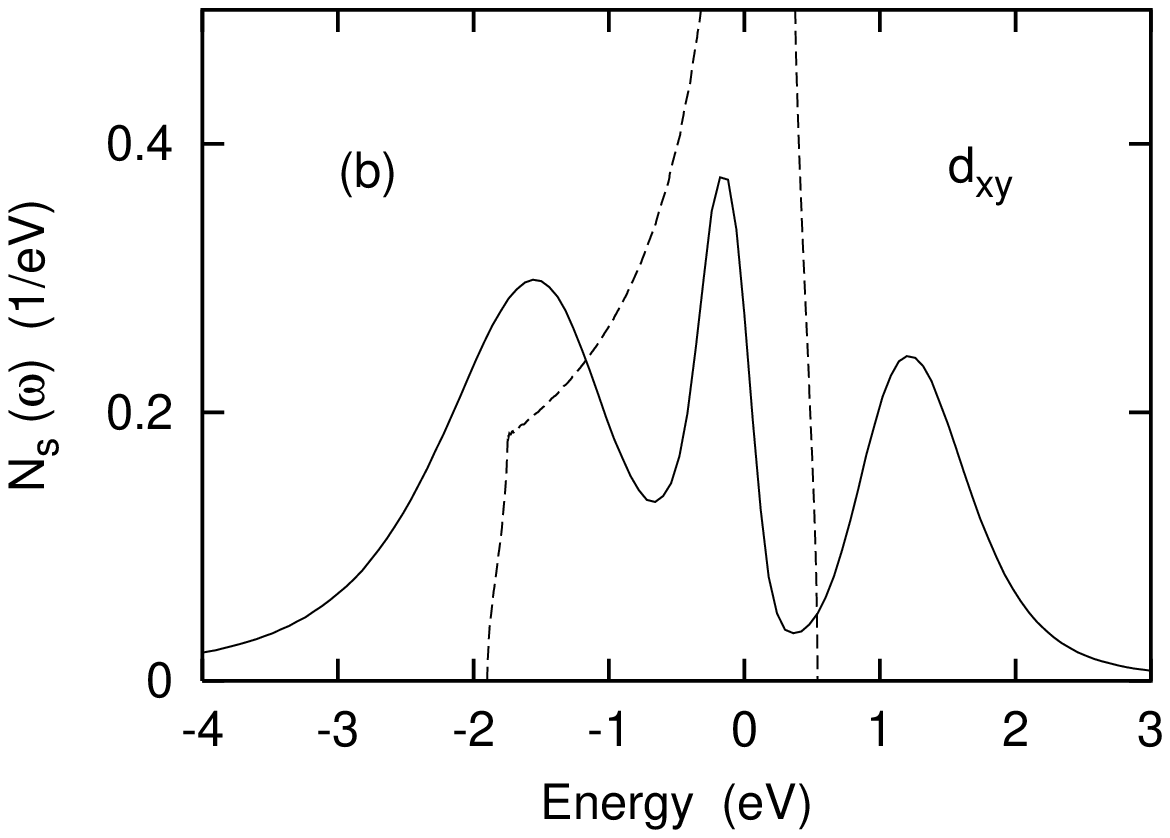}
   \includegraphics[height=7cm,width=4.0cm,angle=-90]{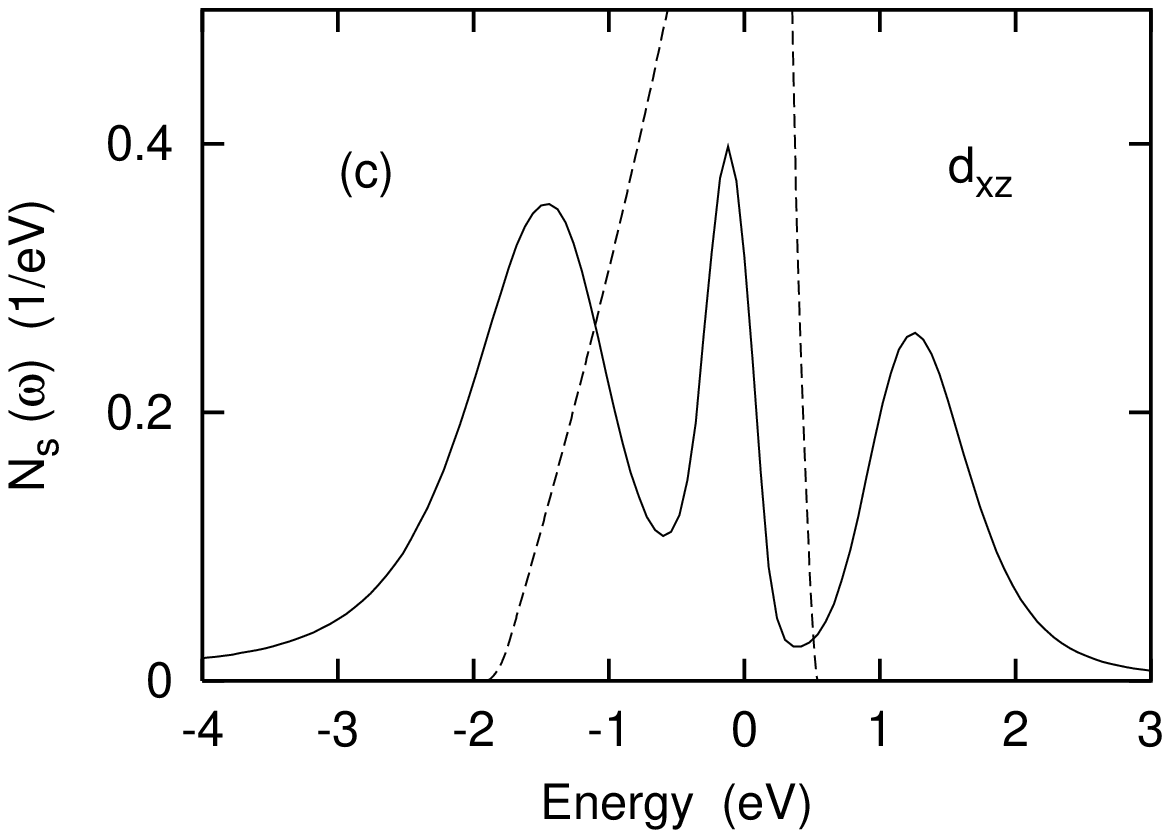}
   \includegraphics[height=7cm,width=4.0cm,angle=-90]{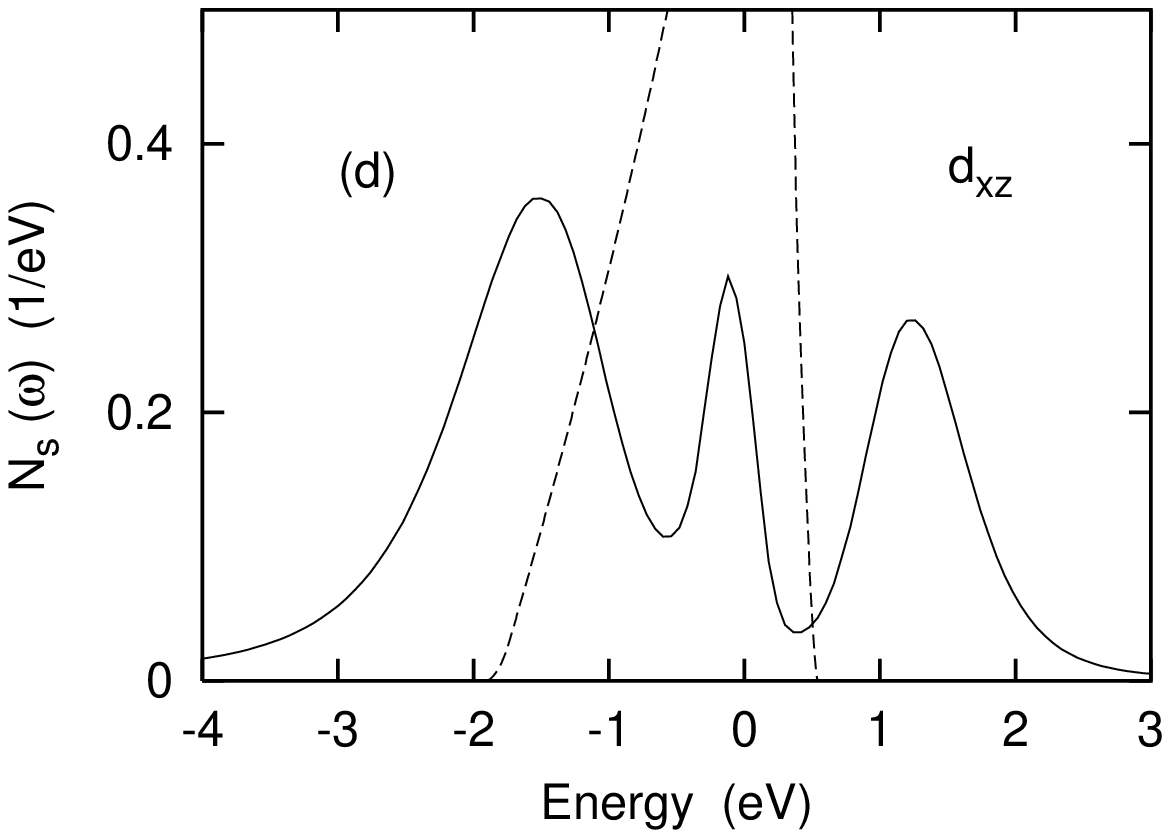}
\end{center}
\caption{
Quasi-particle density of states $N_i(\omega)$ of SrRuO$_3$ ($d^4$) 
derived from DMFT. (a) bulk $t_{2g}$ states; (b) surface $d_{xy}$ states; 
(c) surface $d_{xz,yz}$ states; (d) fictitious isotropic surface 
$d_{xz,yz}$ states (see text).
Solid curves: $U=3.0$\,eV, $J=0.2$\,eV;
dashed curves: bare densities of states (see Fig.~5).
}\end{figure}

To investigate correlations in the bulk and at the surface of SrRuO$_3$ 
we have applied the approach discussed above to the case 
of $d^4$ occupancy of the $t_{2g}$ bands. For the case of cubic 
perovskite structure, i.e., neglecting the tilting of the O octahedra, 
the isotropic bulk density of states shown in Fig.~5 agrees 
qualitatively with results of LAPW calculations\ \cite{mazin}. 
The local density of $d_{xz,yz}$
states in the first layer shows again the characteristic shift
of weight from low and high frequencies to the intermediate 
range closer to $E_F$. The $d_{xy}$ surface density nearly 
coincides with the bulk $t_{2g}$ density. 

Fig.~6 shows the quasi-particle spectra obtained from DMFT-QMC
calculations for $U=3.0$~eV and $J=0.2$~eV. As in the case of  
SrVO$_3$, there is a clear progression of correlation effects
as we go from the isotropic bulk $t_{2g}$ spectra to the $d_{xy}$ 
and $d_{xz,yz}$ surface spectra. The hypothetical isotropic case 
based on the surface $d_{xz,yz}$ density (i.e., equating $\rho_{xy}$ 
with $\rho_{xz,yz}$), shown in Fig.~6\,(d), 
is even more strongly correlated, with less spectral weight of 
the coherent feature near $E_F$ and accordingly larger upper and 
lower Hubbard bands. Again, all bare densities have the same total
width. Thus, the differences between the bulk and surface quasi-particle
spectra are entirely due to the different shapes of the densities.
Note also that, in contrast to the results shown in Fig.~4 for SrVO$_3$,
$N_{b,s}(E_F) < \rho_{b,s}(E_F)$. This deviation is caused by the van 
Hove singularity at $E_F$ and our use of a rather high temperature 
(about 1450~K) in the QMC calculations. 
 
The results shown in Fig.~6 provide a qualitative explanation for the 
experimentally observed shift of spectral weight towards the lower Hubbard 
peak near 1.5~eV binding energy upon lowering the photon energy from 
1254~eV to 100~eV\ \cite{fujioka} and thereby enhancing the surface 
contribution to the photoemission spectra. Of course, as in the case of  
SrVO$_3$, other surface effects, e.g., due to reconstruction, distortion 
of oxygen octahedra, etc., might also play a role and presumably would 
reduce not only the hopping between $d_{xz,yz}$ orbitals of neighboring 
layers but also between $d_{xy}$ states within the surface plane. 
These modifications of the electronic structure would lead to a 
further enhancement of surface correlation effects.   

\subsection{Ca$_x$La$_{1-x}$VO$_3$}

The perovskite series Ca$_x$La$_{1-x}$VO$_3$ is very interesting
in the context of surface versus bulk Coulomb correlations since
LaVO$_3$ ($x=0$) is a Mott insulator ($d^2$ occupancy)\ \cite{solovyev} 
which readily becomes metallic upon increasing the Ca concentration
to $x=0.5$ ($d^{1.5}$ occupancy). Since doping with Ca leads to 
little structural changes and nearly constant Hubbard $U$\ 
\cite{bocquet}, the modifications of the electronic properties 
are almost entirely caused by the degree of band filling. 

To investigate these effects, Maiti {\it et al.}\ \cite{maiti00}
carried out photoemission measurements at Ca concentrations
$x=0.0\ldots0.5$ and using XPS and VUV photon energies to 
distinguish bulk and surface contributions. They observed not 
only the bulk Mott transition at about $x=0.2$, but in addition 
a surface Mott transition at a slightly larger value of $x$.
Thus, these data suggest the coexistence of a metallic bulk with
an insulating surface layer. This finding is in conflict with 
predictions\ \cite{potthoff} for the simple-cubic half-filled 
$s$ band, but it is not clear at present to what extent the 
theoretical results are applicable to multiband systems at
arbitrary occupancies. Moreover, the calculations in Ref.\ 
\cite{potthoff} were done for a linearized version of the DMFT 
in which low- and high-frequency excitations within the $s$ band
are essentially decoupled.      

\begin{figure}[t!]
  \begin{center}
  \includegraphics[width=5.0cm,height=8cm,angle=-90]{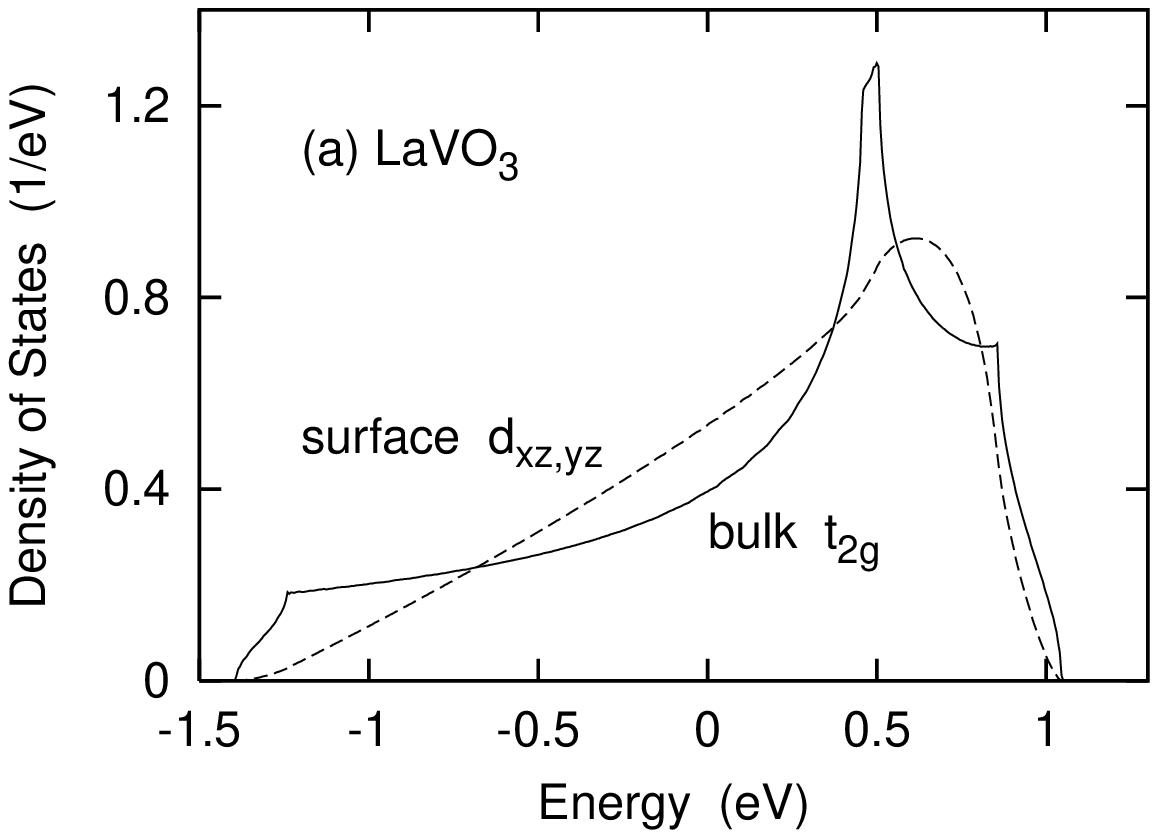}
  \includegraphics[width=5.0cm,height=8cm,angle=-90]{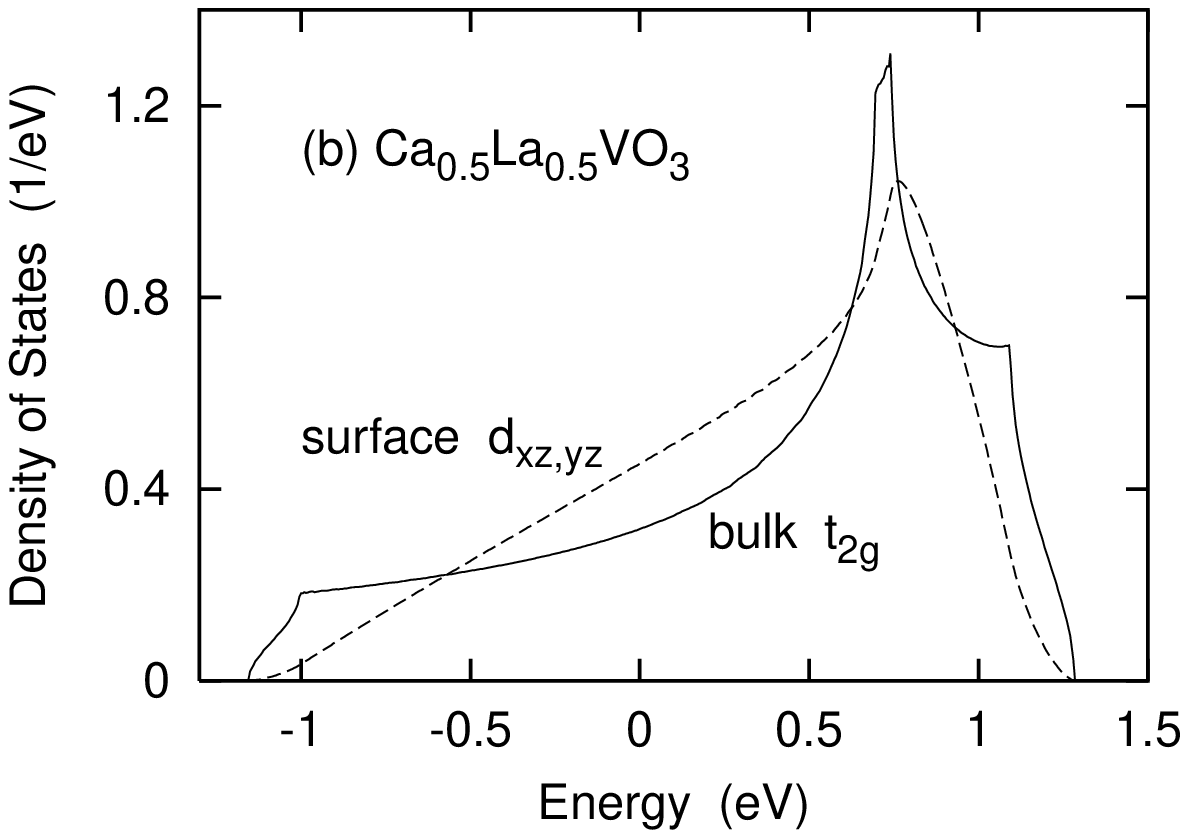}
  \end{center}
\caption{
Solid curves: isotropic bulk density of states of cubic (a) LaVO$_3$ 
($d^2$) and (b) Ca$_{0.5}$La$_{0.5}$VO$_3$ ($d^{1.5}$). Dashed curves: 
local density of $d_{xz,yz}$ states in the first layer ($E_F=0$).
}\end{figure}

We have calculated the quasi-particle spectra of LaVO$_3$ and 
Ca$_{0.5}$La$_{0.5}$VO$_3$ within multi-band QMC-DMFT. The bulk 
density of states is assumed to be similar to the one for cubic 
SrVO$_3$, except for the filling of the $t_{2g}$ bands. As shown 
in Fig.~7, the shape of the $d_{xz,yz}$ density in the surface
layer differs for these two systems since we employ slightly 
different surface potentials in the semi-infinite tight-binding 
calculation to ensure charge neutrality at the surface.
Qualitatively, however, the densities of both systems exhibit a 
similar shift of spectral weight from low and high frequencies 
to the intermediate energy range close to $E_F$. Thus, as a result 
of this effective narrowing of the local density of $d_{xz,yz}$ states 
we can expect a similar enhancement of surface correlation effects 
as discussed above for SrVO$_3$ and SrRuO$_3$.
 
\begin{figure}[t!]
\begin{center}     
    \includegraphics[height=7cm,width=4.0cm,angle=-90]{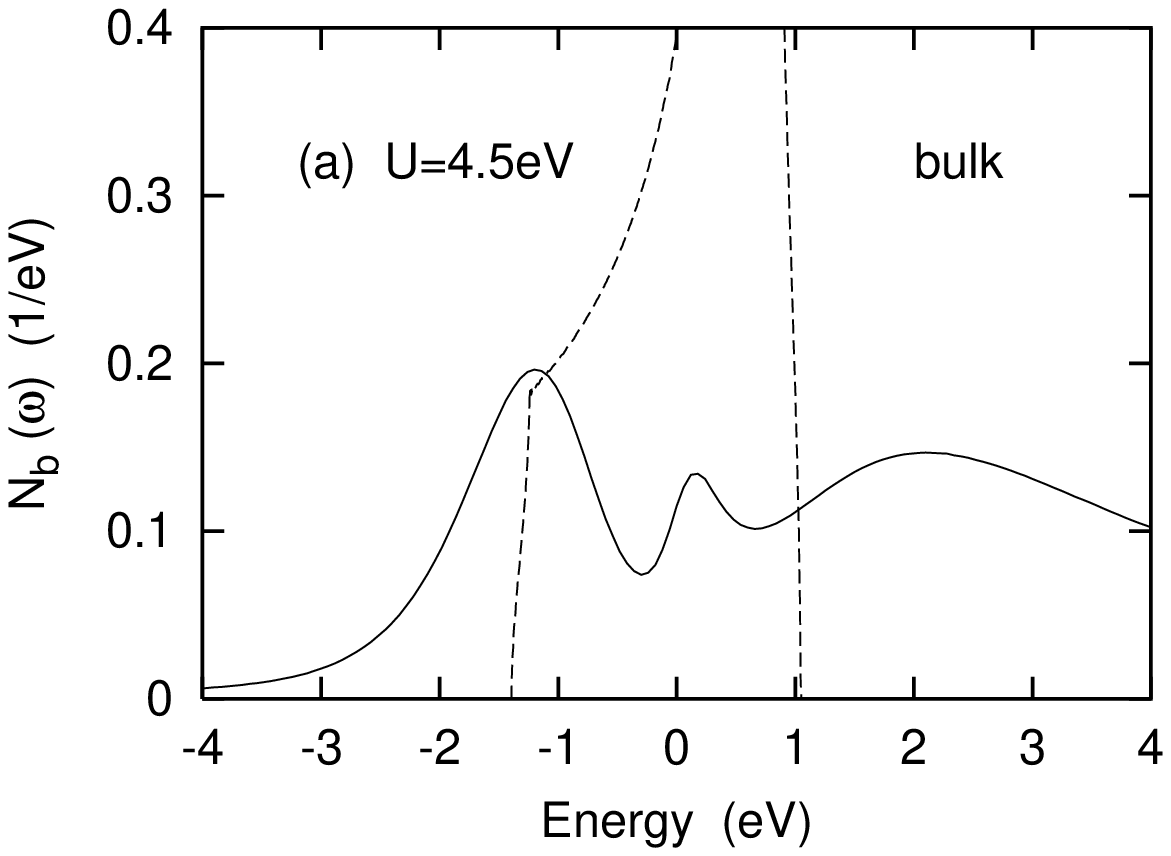}
    \includegraphics[height=7cm,width=4.0cm,angle=-90]{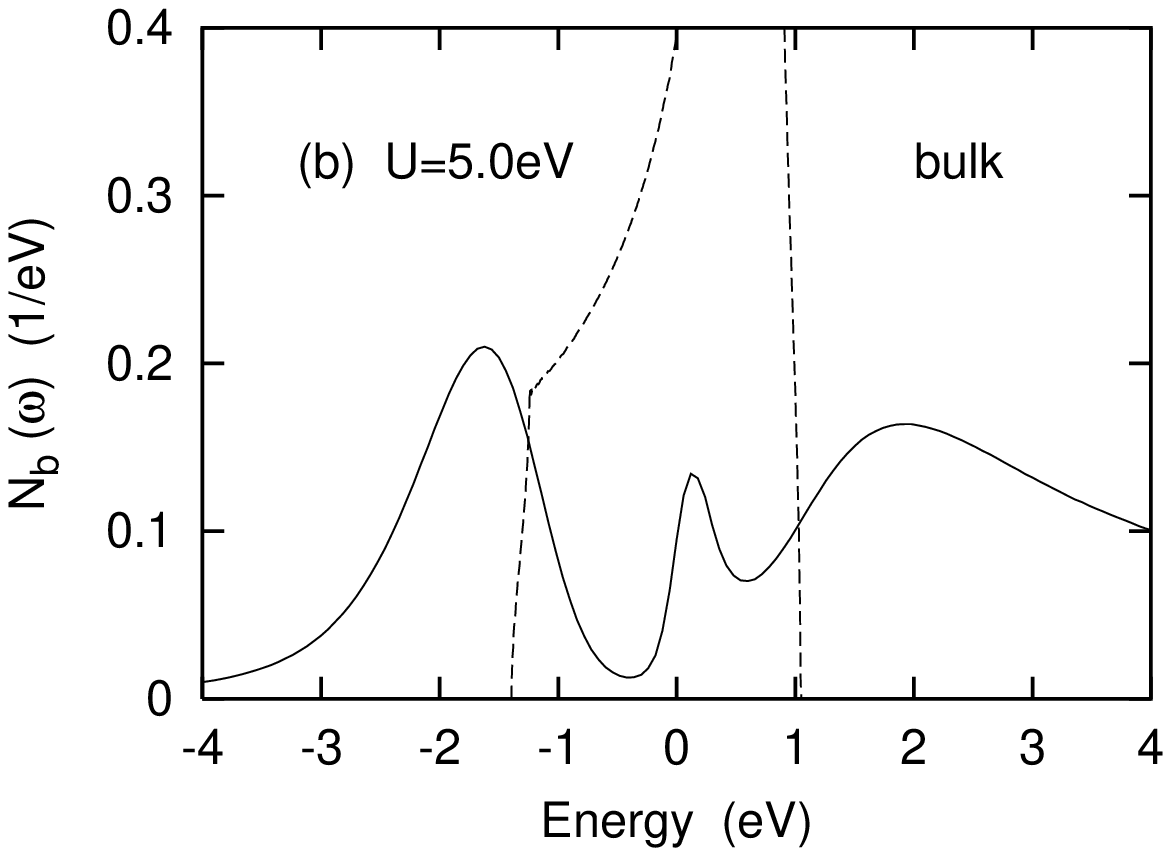}
    \includegraphics[height=7cm,width=4.0cm,angle=-90]{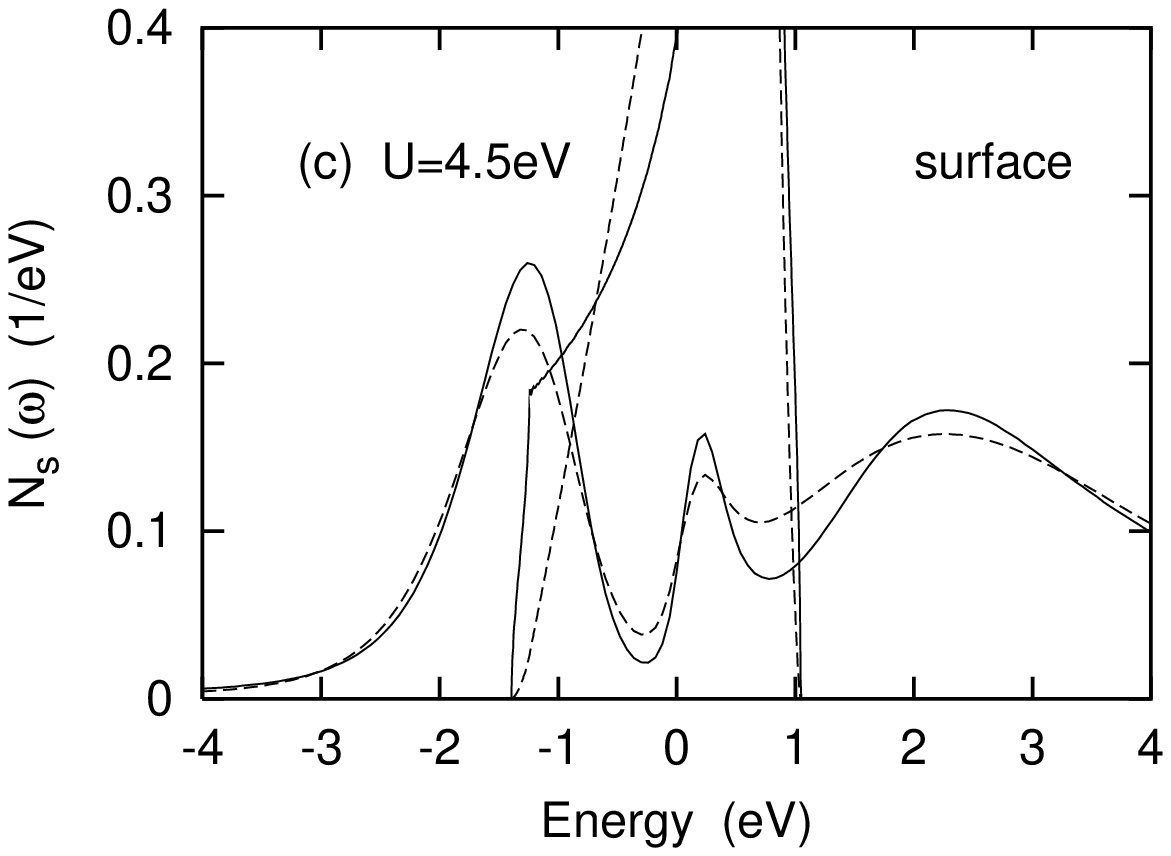}
    \includegraphics[height=7cm,width=4.0cm,angle=-90]{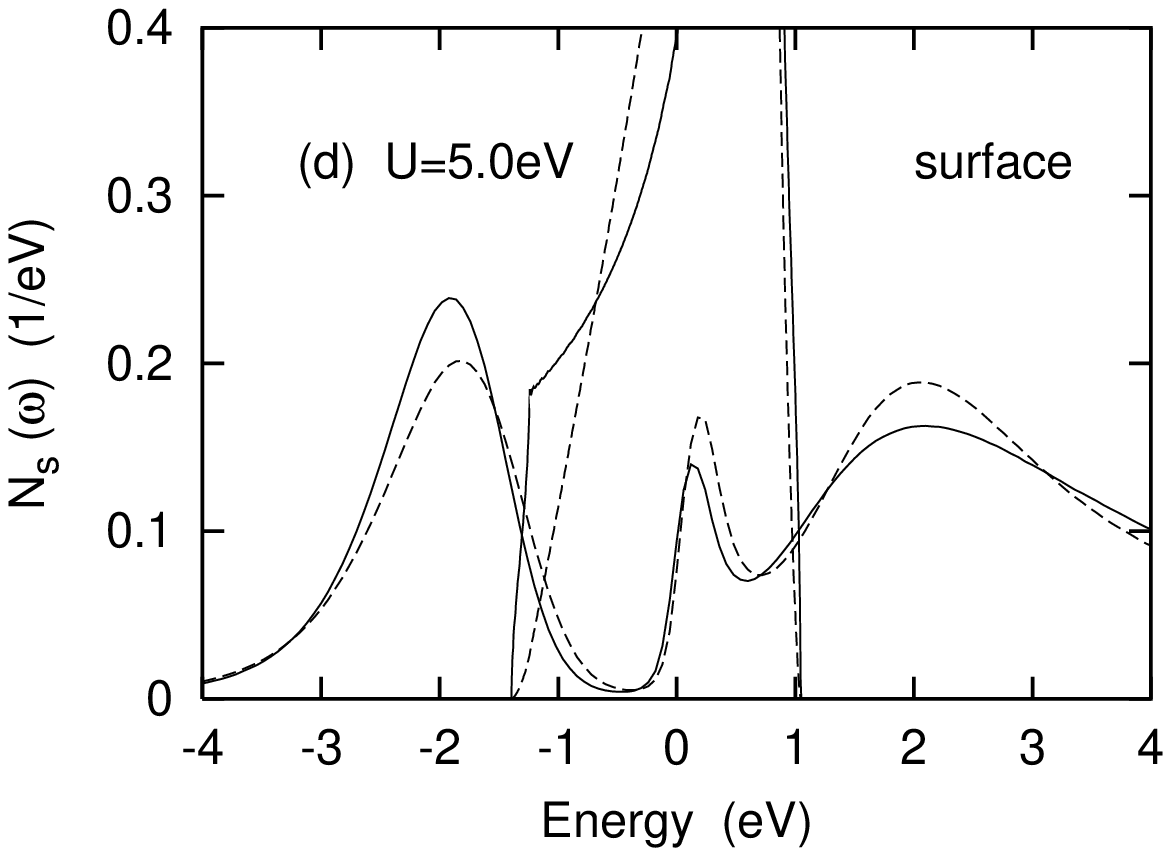}
\end{center}
\caption{
Quasi-particle density of states $N_i(\omega)$ of LaVO$_3$ ($d^2$) 
derived from DMFT. 
(a) and (b): bulk $t_{2g}$ states for $U=4.5$\,eV 
and $U=5.0$\,eV (solid curves; dashed curves: bare densities).
(c) and (d): surface $d_{xy}$ states (solid curves) and 
$d_{xz,yz}$ states (dashed curves) for $U=4.5$\,eV and $U=5.0$\,eV; 
$J=0.7$\,eV.
}\end{figure}

Fig.~8 shows quasi-particle spectra for the bulk and surface 
layer of LaVO$_3$ at Coulomb energies close to the metal-insulator
transition. Since our QMC calculations are carried out at about 1450~K 
($\beta=8$), the resulting coherent peak near $E_F$ obscures the gap 
between upper and lower Hubbard bands which would appear at low
temperatures. Nevertheless, in view of the temperature dependence 
known from one-band systems\ \cite{georges,bulla} 
the results in Fig.~8\,(a) and (b) 
suggest a bulk Mott transition in the range $U=4.5\ldots5.0$~eV.
The surface spectra in  Fig.~8\,(c) and (d) demonstrate that  
correlation effects in the first layer are more pronounced than 
in the bulk, indicating that a gap might form at a slightly lower
value of $U$. However, according to the photoemission experiments\ 
\cite{maiti00} LaVO$_3$ is insulating in the bulk and at the surface. 
Thus, we may conclude that within our model $U$ should be about 5~eV. 

\begin{figure}[t!]
\begin{center}     
    \includegraphics[height=7cm,width=4.0cm,angle=-90]{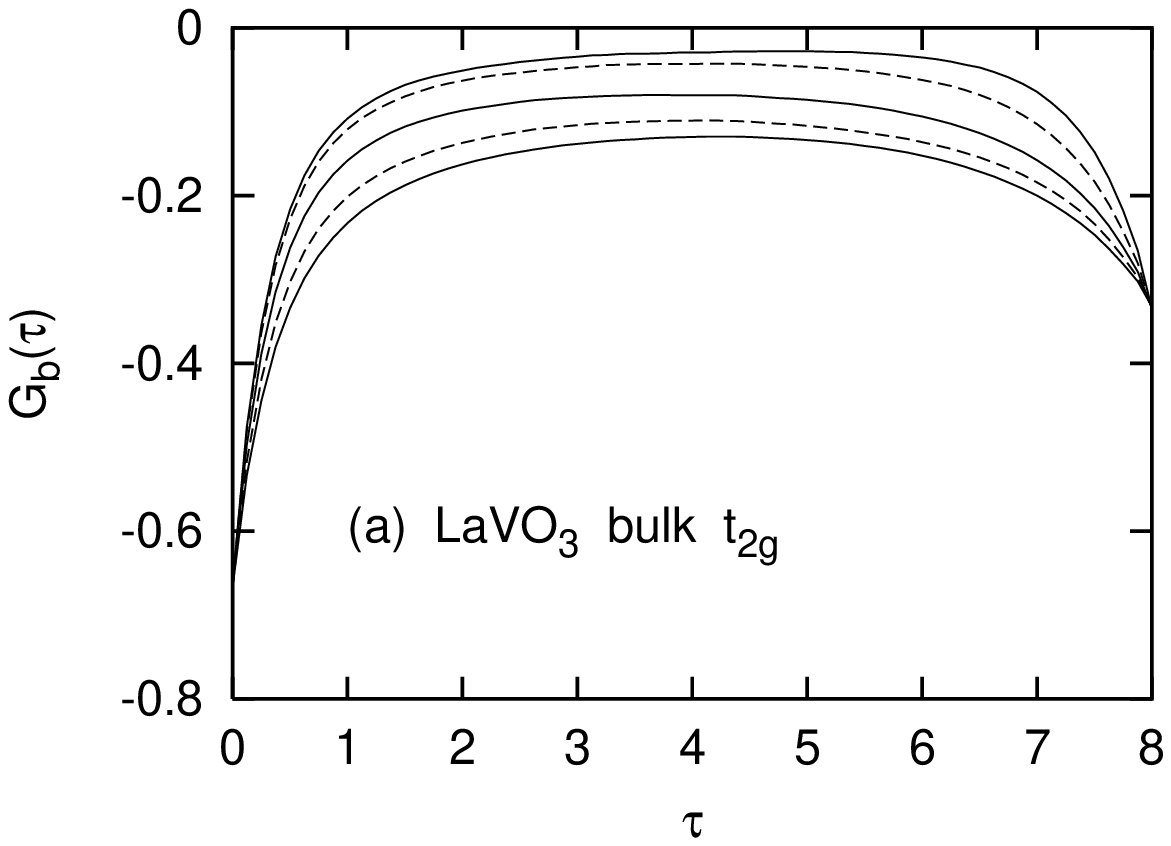}
    \includegraphics[height=7cm,width=4.0cm,angle=-90]{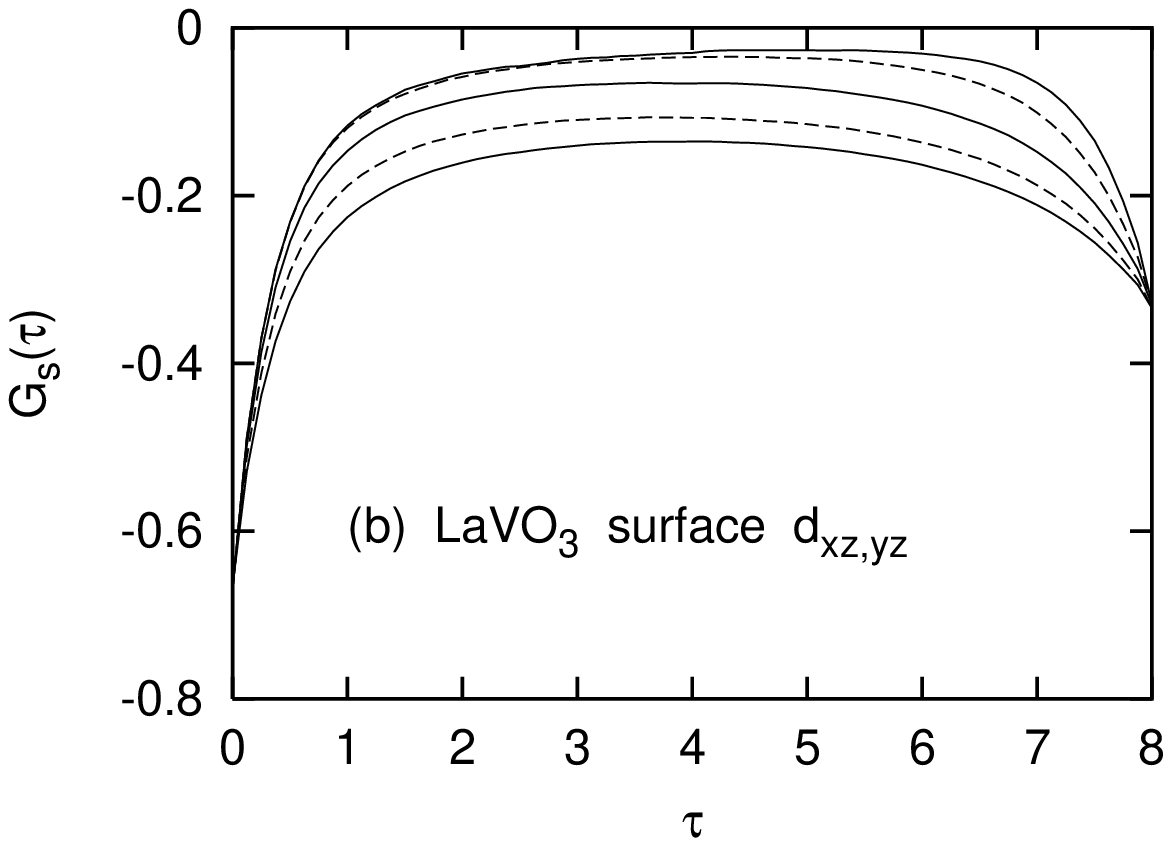}
    \includegraphics[height=7cm,width=4.0cm,angle=-90]{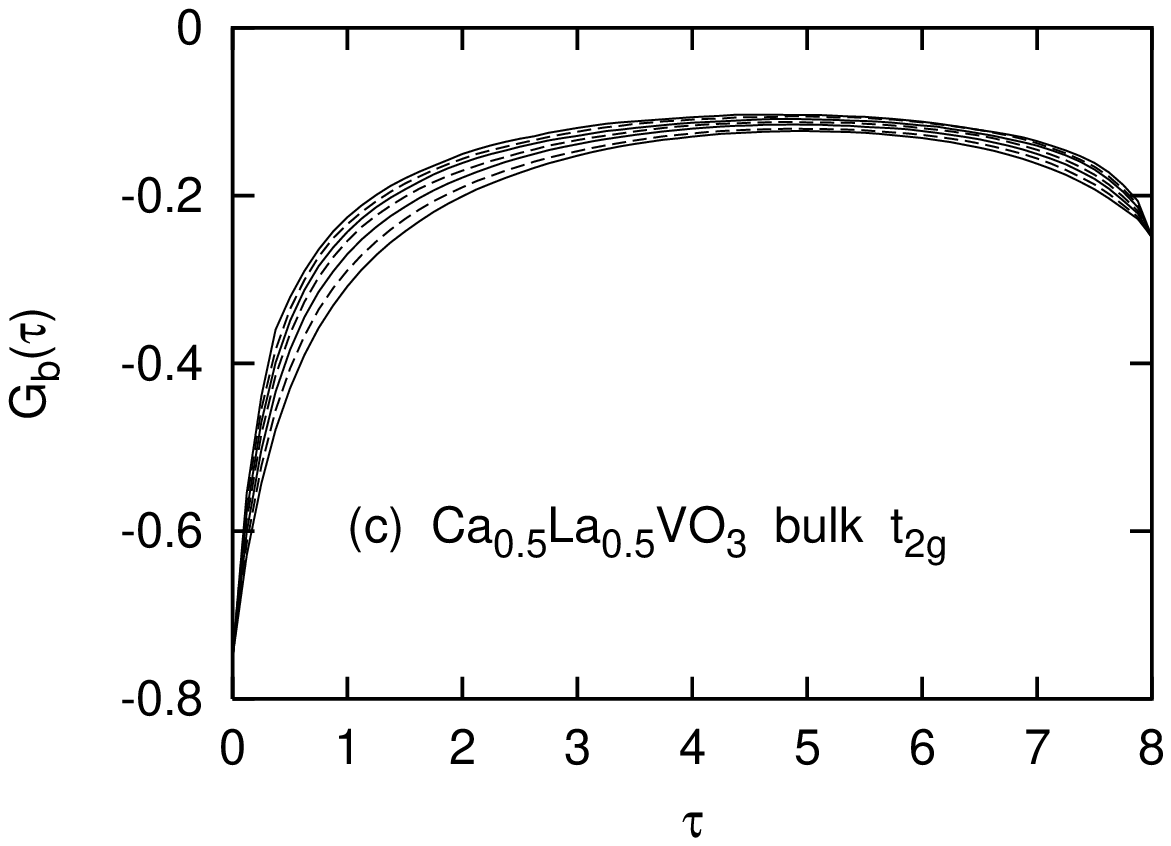}
    \includegraphics[height=7cm,width=4.0cm,angle=-90]{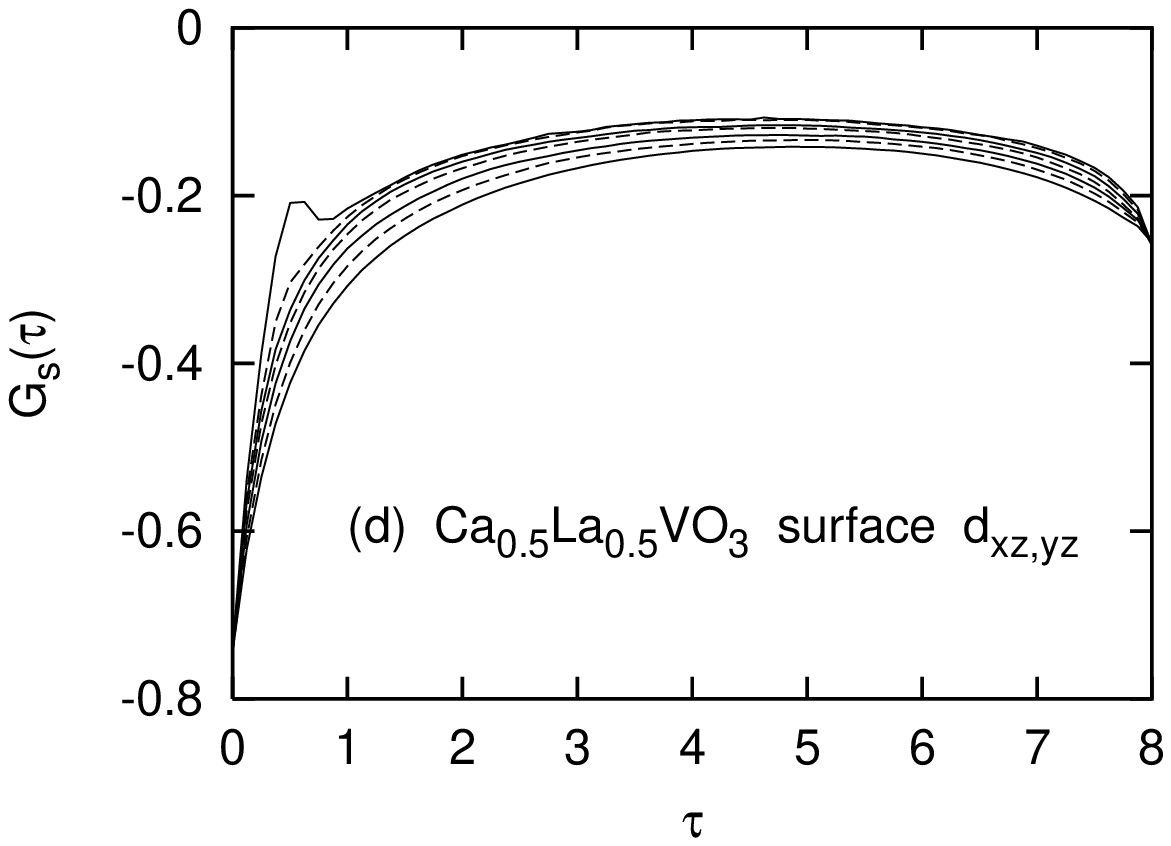}
\end{center}
\caption{
Quasi-particle Green's function as function of imaginary time 
derived from DMFT ($\beta=8$).
(a) bulk $t_{2g}$ states; (b) surface $d_{xz,yz}$ states of LaVO$_3$
($d^2$) for $U=3.0\ldots5.0$~eV in steps of 0.5~eV (from below).
(c) bulk $t_{2g}$ states; (d) surface $d_{xz,yz}$ states of 
Ca$_{0.5}$La$_{0.5}$VO$_3$ ($d^{1.5}$) for $U=3.0\ldots6.0$~eV 
in steps of 0.5~eV (from below). ($J=0.7$~eV). Note that 
$\vert G_i(\beta/2)\vert$ is approximately proportional to 
$N_i(E_F)$.
}\end{figure}

As mentioned above, doping with Ca makes LaVO$_3$ metallic,
leading to rather different excitation spectra. This drastic   
change in electronic properties can be seen clearly in the different
shapes of the imaginary-time Green's function for  LaVO$_3$ and 
Ca$_{0.5}$La$_{0.5}$VO$_3$, as shown in Fig.~9. (The value of 
$G(\tau)$ at $\tau\approx\beta/2$ is representative of the weight 
of the quasi-particle peak at $E_F$, while $G(\beta)$ gives the 
occupancy and $G(0)+G(\beta)=-1$\ \cite{georges}.) In the former 
case, already for $U=4.5\ldots5.0$~eV there is little weight near 
$\tau\approx\beta/2$, indicating the reduction of intensity
near $E_F$. For $x=0.5$, however, even increasing $U$ to 6.0~eV
hardly decreases $G(\beta/2)$. Thus, in the range of Coulomb
energies where LaVO$_3$ is clearly insulating, i.e., $U\approx5.0$~eV,  
Ca$_{0.5}$La$_{0.5}$VO$_3$ shows robust metallic behavior in the 
bulk and at the surface, the surface spectra being slightly more 
correlated than in the bulk. Only for $U=6$~eV the surface 
Green's function for Ca$_{0.5}$La$_{0.5}$VO$_3$ begins to show 
a new feature indicating an instability caused by very strong
local correlations.  

The calculations discussed above suggest that LaVO$_3$ is an insulator 
while Ca$_{0.5}$La$_{0.5}$VO$_3$ is a metal, in agreement experiment. 
To address the interesting question of separate Mott transitions 
in the bulk and at the surface, i.e., the possible coexistence of
metallic bulk and insulating surface properties, it is necessary 
to examine the intermediate range of Ca doping concentrations, 
$0.0<x<0.5$, and to locate the critical region where the bulk has 
just become metallic while the surface is still insulating. This 
issue will be addressed in a subsequent publication.     

\section{conclusion} 

In summary, we have performed multiband QMC-DMFT quasi-particle 
calculations for several perovskite materials using realistic local 
densities of states in the bulk and at the surface. As a result of 
the planar nature of the $t_{2g}$ states, there is an appreciable 
effective narrowing of the local density of states in the first
layer, even though its total width coincides with the one in the
bulk. This band narrowing is entirely a consequence of the reduced
atomic coordination at the surface and does not depend on the 
existence of extra surface states. Typically such states are due
to split-off states below or above the main conduction band and
would not necessarily contribute to the band narrowing. Other
effects, however, which we have not included here, such as tilting 
or distortion of oxygen octahedra in the surface layer, might well
lead to a further reduction of in-plane and/or out-of plane $d-d$ 
hopping parameters and to additional band narrowing. We also note
that the calculations discussed in the present work assume the same
value for the Coulomb and exchange energies at the surface as in 
the bulk. In fact, reduced screening processes close to the
surface could cause an increase of $U$ and thereby make 
surface correlations even more significant.  Moreover, the
termination of the perovskite surfaces, in particular, the charge
state of the transition metal ions, should be investigated. 
   
The main point of our work is that the effective narrowing of the 
surface local density of states leads to an enhancement of   
correlation effects compared to those in the bulk, in agreement 
with photoemission data on various systems. Since photoemission
spectra always involve bulk as well as surface contributions,
it is clearly important to identify the latter in order not to
associate them erroneously with enhanced bulk correlations.  
The pronounced two-dimensional characteristics of $d$ states 
near $E_F$ is a common feature of many transition metal oxides. 
Thus, the surface-induced enhancement of correlation effects 
discussed in the present work should be a phenomenon 
observable in many materials.   

We finally remark that all of the systems discussed in the present
work have very similar, albeit highly asymmetric bulk densities of 
states. They differ primarily by the degree of band filling, i.e., by 
the position of the Fermi level with respect to the density of states
features. Also, the surface densities of the out-of-plane subbands
of the various systems are qualitatively similar. Nevertheless, even
though the quasi-particle spectra for sufficiently large on-site
Coulomb energies all exhibit the familiar three-peak structure, with 
a quasi-particle peak at $E_F$ and lower and upper Hubbard satellites, 
the positions, widths, asymmetric shapes and relative weights 
of these three peaks vary considerably for the different materials. 
To describe these variations 
which evidently reflect the original densities and band fillings,
one would need at least 11 parameters (the total area integrates
to unity), suggesting that even in the strong-correlation limit
the spectra reveal a high degree of system-specific information 
concerning their electronic structure. Obviously, far greater 
complexity can be expected for non-cubic and magnetic materials. 
The adequate description of this complexity which lies at the root
of many fascinating material properties underlines the tremendous
progress that has been achieved by combining density functional
theory and quantum impurity methods via the Dynamical Mean Field
Theory.

\bigskip
Acknowledgments:

I like to thank A. Bringer for useful discussions and A.I. Lichtenstein 
for the QMC-DMFT code.

\end{document}